\documentclass[11pt,a4paper,twoside,groupcitations]{article}
\usepackage[T1]{fontenc}
\usepackage[ansinew]{inputenc}
\usepackage[english]{babel}
\usepackage{amsfonts}
\usepackage{amsmath}
\usepackage{array}
\usepackage{amsthm}
\usepackage{amssymb}
\usepackage{graphicx}
\usepackage{subfigure}
\usepackage{braket}
\usepackage{eucal}
\usepackage{verbatim}
\usepackage[table]{xcolor}
\usepackage{caption}
\usepackage{cite}
\usepackage{multicol}
\usepackage{tikz}
\usetikzlibrary{positioning,arrows}
\usetikzlibrary{decorations.pathmorphing}
\usetikzlibrary{decorations.markings}
\usetikzlibrary{calc,decorations.markings}
\usetikzlibrary{arrows,shapes}
\usetikzlibrary{matrix,arrows}
\usepackage{pgfplots}
\usepackage{xparse}
\definecolor{jade}{HTML}{00A86B}
\newcommand{\be}{\begin{eqnarray}}
\newcommand{\ee}{\end{eqnarray}}

\renewcommand{\d}{\mbox{${\rm d}$}} 
\newcommand{\lp}{\ell_{\rm p}}
\newcommand{\mpl}{m_{\rm p}}
\newcommand{\gn}{G_{\rm N}}
\newcommand{\rh}{r_{\rm H}}
\newcommand{\Rh}{R_{\rm H}}

\title{\bf Bootstrapping Newtonian Gravity}
\author{Roberto~Casadio$^{ab}$\thanks{E-mail: casadio@bo.infn.it},
$\ $
Michele~Lenzi$^{ab}$\thanks{E-mail: michele.lenzi@studio.unibo.it}
$\ $
and
Octavian Micu$^c$\thanks{E-mail: octavian.micu@spacescience.ro}
\\
\\
$^a${\em Dipartimento di Fisica e Astronomia, Universit\`a di Bologna}
\\
{\em via Irnerio~46, 40126 Bologna, Italy}
\\
\\
$^b${\em I.N.F.N., Sezione di Bologna, I.S.~FLAG}
\\
{\em viale B.~Pichat~6/2, 40127 Bologna, Italy}
\\
\\
{\em $^b$Institute of Space Science, Bucharest, Romania}
\\
{\em P.O. Box MG-23, RO-077125 Bucharest-Magurele, Romania}
}
\begin{document}
\maketitle
\begin{abstract}
A non-linear equation obtained by adding gravitational self-interaction terms to the Poisson equation for
Newtonian gravity is here employed in order to analyse a static spherically symmetric homogeneous
compact source of given proper mass and radius and the outer vacuum.
The main feature of this picture is that, although the freedom of shifting the potential by an arbitrary
constant is of course lost, the solutions remain qualitatively very close to the Newtonian behaviour.
We also notice that the negative gravitational potential energy is smaller than the
proper mass for sources with small compactness, but for sources that should form black holes
according to General Relativity, the gravitational potential energy becomes of the same order of
magnitude of the proper mass, or even larger.
Moreover, the pressure overcomes the energy density for large values of the compactness,
but it remains finite for finite compactness, hence there exists no Buchdahl limit.
This classical description is meant to serve as the starting point for investigating quantum
features of (near) black hole configurations within the corpuscular picture of gravity in future
developments.
\par
\null
\par
\noindent
\textit{PACS - 04.70.Dy, 04.70.-s, 04.60.-m}
\end{abstract}
\section{Introduction}
\setcounter{equation}{0}
\label{Sintro}
It is well-known that Newtonian gravity can be recovered from General Relativity in the
weak field and non-relativistic regime~\cite{weinberg}.
It was also shown long ago that General Relativity is in turn the simplest among the 
(diffeomorphism invariant) consistent completions of the linear Newtonian gravity with the graviton
self-coupling~\cite{deser}. 
This can be more easily demostrated in the first order Palatini formalism, in which the metric $g$
and the connection $\Gamma$ are initially treated as independent variables. 
The relation that makes $\Gamma$ compatible with the metric is then derived and 
contains the inverse of $g$, which makes it apparent why this completion is essentially
non-perturbative.
Later on, it was shown~\cite{wald} that there in fact exists a larger class of Lorentz invariant
theories for interacting massless gravitons which are consistent with quantum physics
(see also Refs.~\cite{unruh,hertzberg,bai}).
\par
In the present work we wish to tackle a much more modest task than recovering classical General Relativity
and all of its fundamental symmetries, namely we will study an effective equation for the gravitational potential
of a static source which contains a gravitational self-interaction term besides the usual coupling
with the matter density.
Following an idea from Ref.~\cite{Casadio:2016zpl}, this equation was derived in details
from a Fierz-Pauli Lagrangian in Ref.~\cite{Casadio:2017cdv}, and it can therefore be viewed
as stemming from the truncation of the fully relativistic theory at some ``post-Newtonian'' order
(for the standard post-Newtonian formalism, see Ref.~\cite{PPNlr}).
The main motivation for this study is provided by the quantum model of corpuscular black holes~\cite{DvaliGomez},
in which the constituents of black holes are assumed to be gravitons marginally bound in their own
gravitational potential well.
The typical size of this well is given by the characteristic Compton-de~Broglie wavelength
$\lambda_{\rm G}\sim \Rh$, where~\footnote{We shall mostly use units with $c=1$ and the Newton
constant $\gn=\lp/\mpl$, where $\lp$ is the Planck length and $\mpl$ the Planck mass
(so that $\hbar=\lp\,\mpl$).}
\be
\Rh=2\,\gn\,M
\label{Rh}
\ee
is the Schwarzschild (gravitational) radius of the black hole of mass $M$, and whose depth is proportional to
the very large number 
\be
N_{\rm G}
\sim 
\frac{M^2}{\mpl^2}
\sim
\frac{\Rh^2}{\lp^2}
\label{Ng}
\ee
of soft quanta in this condensate.
When the contribution of gravitons is related to the necessary presence of ordinary
matter, the picture appears connected to the post-Newtonian approximation~\cite{Casadio:2016zpl}.
This can be seen by considering that the (negative) gravitational energy of a source of mass $M$
localised inside a sphere of radius $R$ is given by
\be
U_{\rm N}
\sim
M\, V_{\rm N}
\sim
-\frac{\gn\,M^2}{R}
\ ,
\ee
where $V_{\rm N}\sim -{\gn\,M}/{R}$ is the (negative) Newtonian potential.
This potential can be represented by the expectation value of a scalar field over a coherent state $\ket{g}$,
whose normalisation then yields the graviton number~\eqref{Ng}, which reproduces Bekenstein's area
law~\cite{bekenstein}.
In addition to that, assuming most gravitons have the same wave-length $\lambda_{\rm G}$,
the (negative) energy of each single graviton is correspondingly given by
\be
\epsilon_{\rm G}
\sim
\frac{U_{\rm N}}{N_{\rm G}}
\sim
-\frac{\mpl\,\lp}{R} 
\ ,
\ee
which yields the typical Compton-de~Broglie length $\lambda_{\rm G}\sim R$.
The graviton self-interaction energy hence reproduces the (positive) post-Newtonian energy,
\be
U_{\rm GG}(R)
\sim
N_{\rm G} \, \epsilon_{\rm G} \, V_{\rm N}
\sim
\frac{\gn^2\,M^3}{R^2}
\ ,
\ee
and the fact that gravitons in a black hole are marginally bound is reflected by the ``maximal
packing'' condition~\cite{DvaliGomez}, which roughly reads $U_{\rm N}+U_{\rm GG}\simeq 0$
for $R=\Rh$~\cite{Casadio:2016zpl,Casadio:2017cdv,dadhich2}.
\par
The above quantum picture was previously tested by studying small (post-Newtonian) perturbations around
the Newtonian potential in Ref.~\cite{Casadio:2017cdv}.
However, since the post-Newtonian correction $V_{\rm PN}\sim 1/r^2$ is positive and grows faster than
the Newtonian potential near the surface of the source, one is allowed to consider only matter sources with
radius $R\gg\Rh$ for this approximation to hold.
This consistency condition clearly excludes the possibility to study very compact matter sources and, in particular,
those on the verge of forming a black hole, that is with $R\simeq \Rh$.
For the ultimate purpose of including such cases, we shall here study the non-linear equation of the effective theory
derived in Ref.~\cite{Casadio:2017cdv} at face value, without requiring that the corrections it introduces with respect
to the Newtonian potential remain small.
We shall nonetheless show that the qualitative behaviour of the complete solutions to our non-linear equation
resembles rather closely the Newtonian counterpart.
This result, which essentially stems from including a gravitational self-interaction in the Poisson equation, 
is what we call ``bootstrapping'' the Newtonian gravity.
\par
The paper is organised as follows:
in the next Section we briefly review the derivation of the effective equation for the gravitational potential
obtained by including a potential self-interaction;
in Section~\ref{S:vacuum}, we find the exact solution in the vacuum and, in Section~\ref{S:solution},
we analyse the case of a homogenous spherical source;
a possible connection with the corpuscular model and future perspectives are then discussed
in the concluding Section~\ref{S:conc}.
\section{Effective theory for the gravitational potential}
\label{S:action}
\setcounter{equation}{0}
We shall start by briefly recalling the main steps in the derivation of the non-linear equation which reproduces
the Newtonian potential to leading order and includes the effects of gravitational self-interaction obtained in
Ref.~\cite{Casadio:2017cdv}.
First of all, one assumes the local curvature is small, so that the metric can be written
as $g_{\mu\nu}=\eta_{\mu\nu}+h_{\mu\nu}$, where $\eta_{\mu\nu}$ is the flat
Minkowski metric with signature $(-,+,+,+)$ and $|h_{\mu\nu}|\ll 1$.
In addition to this weak field limit, we must assume that all matter in the system moves with
a characteristic velocity much slower than the speed of light in the (implicitly) chosen
reference frame $x^\mu=(t,{\bf x})$.
In fact, we shall just consider (static) spherically symmetric systems, so that $\rho=\rho(r)$ and
the only relevant component of the metric is therefore $h_{00}(r)\equiv -2\,V(r)$.
In this approximation, the Einstein-Hilbert Lagrangian with matter reduces to
\be
L_{\rm N}[V]
\simeq
-4\,\pi
\int_0^\infty
r^2 \,\d r
\left[
\frac{\left(V'\right)^2}{8\,\pi\,\gn}
+\rho\,V
\right]
\ ,
\label{LagrNewt}
\ee
where $f'\equiv\d f/\d r$.
The corresponding equation of motion is just the Poisson equation,
\be
r^{-2}\left(r^2\,V'\right)'
\equiv
\triangle V
=
4\,\pi\,\gn\,\rho
\ ,
\label{EOMn}
\ee
for the Newtonian potential $V=V_{\rm N}$.
\par
In order to go beyond the Newtonian approximation, we modify the latter functional by adding a
non-linear term.
This term can be obtained by noting that the Hamiltonian
\be
H_{\rm N}[V]
=
-L_{\rm N}[V]
=
4\,\pi
\int_0^\infty
r^2\,\d r
\left(
-\frac{V\,\triangle V}{8\,\pi\,\gn}
+\rho\,V
\right)
\ ,
\label{NewtHam}
\ee 
computed on-shell by means of Eq.~\eqref{EOMn}, yields the Newtonian potential energy
\be
U_{\rm N}(r)
&\!\!=\!\!&
2\,\pi
\int_0^r 
{\bar r}^2\,\d {\bar r}
\,\rho(\bar r)\, V(\bar r)
\nonumber
\\
&\!\!=\!\!&
\frac{1}{2\,\gn}
\int_0^r 
{\bar r}^2 \,\d {\bar r}\,
V(\bar r)\,\triangle V(\bar r)
\notag
\\
&\!\!=\!\!&
-\frac{1}{2\,\gn}\,
\int_0^r 
{\bar r}^2 \,\d {\bar r}\,
\left[V'(\bar r) \right]^2
\ ,
\label{Unn}
\ee
where we used Eq.~\eqref{EOMn} and then integrated by parts discarding boundary terms.
One can view the above $U_{\rm N}$ as given by the interaction of the matter distribution enclosed in a sphere
of radius $r$ with the gravitational field.
Following Ref.~\cite{Casadio:2016zpl} (see also Ref.~\cite{dadhich1}), we then define a self-gravitational
source $J_V$ proportional to the gravitational energy $U_{\rm N}$ per unit volume, that is~\cite{Casadio:2017cdv}  
\be
J_V(r)
=
\frac{4}{4\,\pi\, r^2}\,\frac{\d}{\d r} U_{\rm N}(r)
=
-\frac{1}{2\,\pi\,\gn}
\left[ V'(r) \right]^2
\ .
\label{JV}
\ee
Upon including this new source term, and the analogous higher order term $J_\rho=-2\,V^2$
which couples with the matter source, we obtain the total Lagrangian~\cite{Casadio:2017cdv} 
\be
L[V]
&\!\!=\!\!&
L_{\rm N}[V]
-4\,\pi
\int_0^\infty
r^2\,\d r
\left(
q_\Phi\,J_V\,V
+
q_\Phi\,J_\rho\,\rho
\right)
\nonumber
\\
&\!\!=\!\!&
-4\,\pi
\int_0^\infty
r^2\,\d r
\left[
\frac{\left(V'\right)^2}{8\,\pi\,\gn}
\left(1-4\,q_\Phi\,V\right)
+\rho\,V\left(1-2\,q_\Phi\,V\right)
\right]
\ .
\label{LagrV}
\ee
Finally, the Euler-Lagrange equation for $V$ is thus given by
\be
\left(1-4\,q_\Phi\,V\right)\left(\triangle V-4\,\pi\,\gn\,\rho\right)
=
2\,q_\Phi
\left(V'\right)^2
\ ,
\label{EOMV}
\ee
where $q_\Phi$ is a self-coupling parameter. 
In particular, it was shown in Ref.~\cite{Casadio:2017cdv} that the first post-Newtonian 
expansion of the Schwarzschild metric is recovered for $q_\Phi=1$, and we shall
therefore assume this value in the following (we will briefly come back to this point in
the Conclusions).
\par
In the next sections, we shall analyse Eq.~\eqref{EOMV} as an effective description of 
the static gravitational field $V$ generated by a static source of density $\rho$ in flat space-time.
In other words, we abandon, or disregard, its geometric origin given by the Einstein-Hilbert
action and proceed by assuming there exists a reference frame in which the motion of
test particles are described by Newton's law with a potential that solves Eq.~\eqref{EOMV}.
Before we try and solve this equation, it is then important to note that the freedom to shift the
Newtonian potential by an arbitrary constant, say $V_0$, is now lost in general.
In fact, if $V_{\rm c}$ solves Eq.~\eqref{EOMV}, for any $\bar V=V_{\rm c}+V_0$
one finds
\be
\left[1-4 \left(V_c+V_0\right)\right]
\left(\triangle V_c-4\,\pi\,\gn\,\rho\right)
-2 \left(V_{\rm c}'\right)^2
=
-4\,V_0
\left(\triangle V_{\rm c}-4\,\pi\,\gn\,\rho\right)
\ ,
\label{noV0}
\ee
which therefore means that $\bar V$ would still be a solution only if $V_{\rm c}=V_{\rm N}$.
This property clearly parallels General Relativity.
We however note that Eq.~\eqref{LagrV} does not yet contain the pressure as a source,
which will have to be added in order to ensure energy conservation in general, as we will 
comment more extensively in Section~\ref{S:conc}.
\par
On a qualitative ground, one might expect that the term in the r.h.s.~of Eq.~\eqref{EOMV} becomes less
important for $V_{\rm c}$ negative and large, and one approximately recovers the Poisson
equation~\eqref{EOMn} in this case.
We shall in fact see that the solution to Eq.~\eqref{EOMV} can be conveniently expressed
as a (somehow small) perturbation about the Newtonian potential where $\rho\not = 0$.
On the other hand, in the vacuum $\rho=0$ and the effect of the new gravitational self-coupling
in Eq.~\eqref{EOMV} leads to (possibly) significant departures from the Newtonian behaviour.
\section{Spherical vacuum solution}
\label{S:vacuum}
\setcounter{equation}{0}
In the vacuum, where $\rho=0$, Eq.~\eqref{EOMV} with $q_\Phi=1$ reads
\be
\triangle V
=
\frac{2 \left(V'\right)^2}{1-4\,V}
\ ,
\label{EOMV0}
\ee
which is exactly solved by
\be
V_{\rm c}
=
\frac{1}{4}
\left[1-c_1\left(1+\frac{c_2}{r}\right)^{2/3}
\right]
\ .
\label{sol0cc}
\ee
where $c_1$ and $c_2$ are integration constants, and we note again that that one cannot
shift $V_{\rm c}$ by an arbitrary constant $V_0$.
\begin{figure}[t]
\centering
\includegraphics[width=8cm]{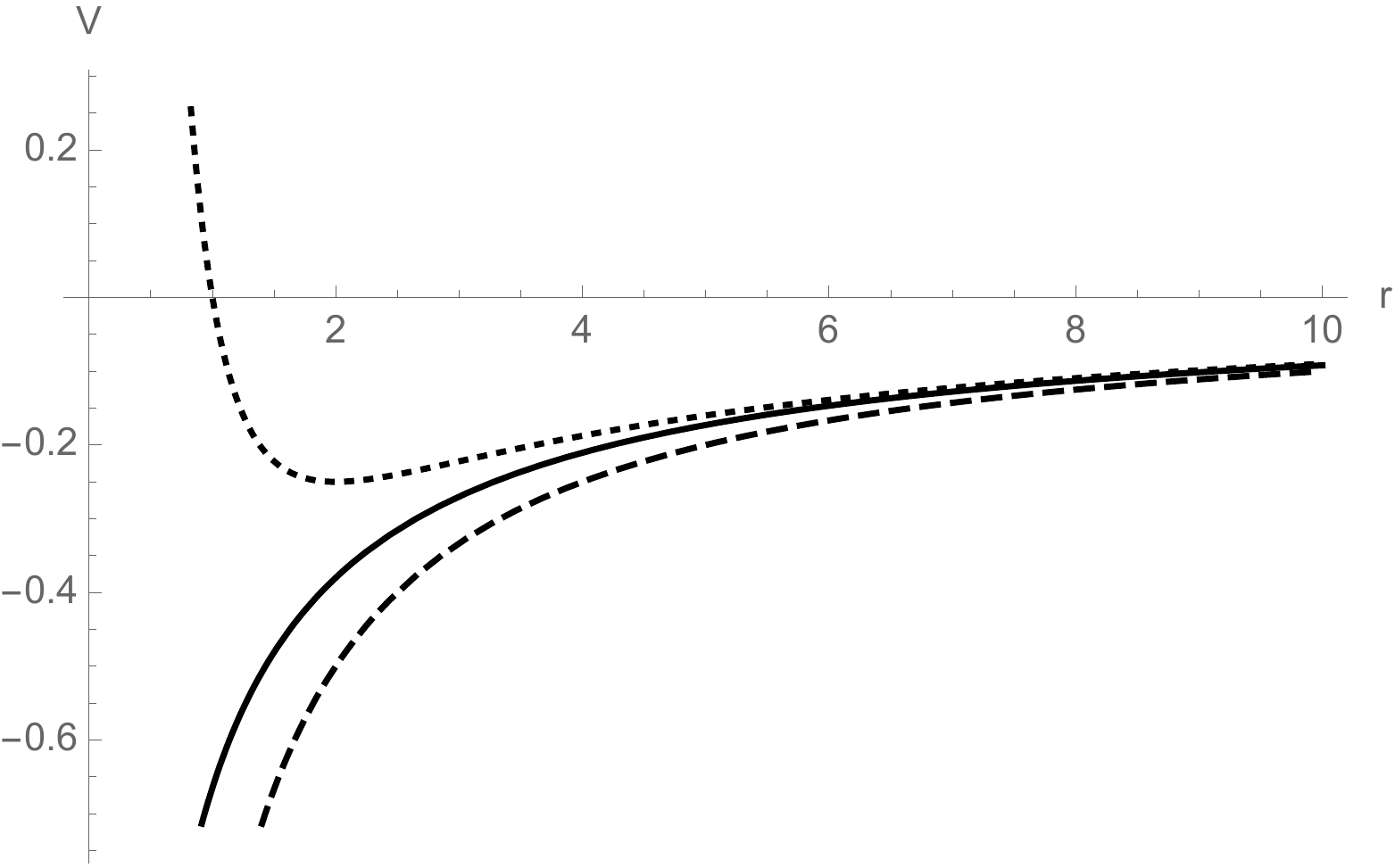}
$\ $
\includegraphics[width=8cm]{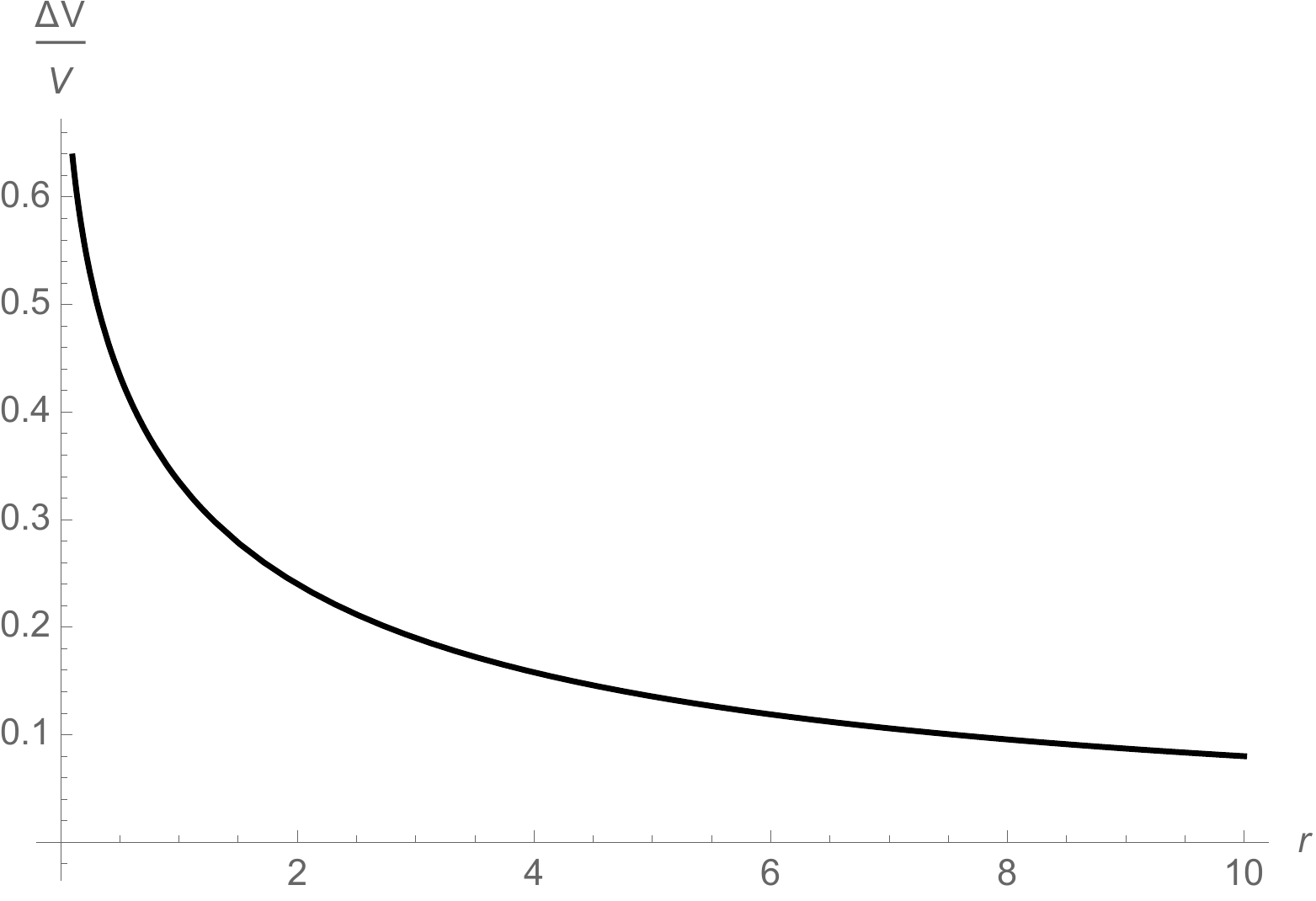}
\caption{Left panel:
potential $V_{\rm c}$ (solid line) {\em vs\/} Newtonian potential (dashed line)
{\em vs\/} order $\gn^2$ expansion of $V_{\rm c}$ (dotted line) for $r>0$.
Right panel: 
relative difference $(V_{\rm N}-V_{\rm c})/V_{\rm N}$ for $r>0$
(all quantities are in units of $\gn\,M$).
}
\label{V0}
\end{figure}
\par
The two integration constants can still be fixed by requiring the expected
Newtonian behaviour in terms of the ADM mass~\cite{adm} $M$ for large $r$. 
One must then have $c_1=1$ and $c_2=6\,\gn\,M$, which yield
\be
V_{\rm c}
=
\frac{1}{4}
\left[
1-\left(1+\frac{6\,\gn\,M}{r}\right)^{2/3}
\right]
\ .
\label{sol0}
\ee
The large $r$ expansion now reads
\be
V_{\rm c}
\underset{r\to\infty}{\simeq}
-\frac{\gn\,M}{r}
+\frac{\gn^2\,M^2}{r^2}
-\frac{8\,\gn^3\,M^3}{3\,r^3}
\ ,
\label{Vlarge}
\ee
and contains the expected post-Newtonian term $V_{\rm PN}$ of order $\gn^2$ without any further
assumptions~\cite{Casadio:2017cdv}.
Moreover, unlike its truncation at order $\gn^2$, the above $V_{\rm c}$ tracks the Newtonian 
solution for all values $r>0$ (see Fig.~\ref{V0}).
In particular, $V_{\rm c}$ remains (increasingly) negative like $V_{\rm N}$, but diverges
slower than $V_{\rm N}$ for $r\to 0$, since
\be
\frac{V_{\rm c}}{V_{\rm N}}
\underset{r\to 0}{\sim}
\left(\frac{r}{\gn\,M}\right)^{1/3}
\ ,
\ee
as is also displayed by the relative difference in the right panel of Fig.~\ref{V0}.
This shows that the added source in the r.h.s.~of Eq.~\eqref{EOMV0} acts as a (partial) regulator.
In Fig.~\ref{V0}, we also plot the large $r$ expansion~\eqref{Vlarge} up to the first post-Newtonian
approximation of order $\gn^2$, which describes a repulsive force for $r<2\,\gn\,M$.
Of course, gravity is attractive and one must view this result as indicating the post-Newtonian
expansion fails at such short distances.
\section{Homogeneous ball in vacuum}
\label{S:solution}
\setcounter{equation}{0}
Since we are interested in compact sources, we will consider the simplest case in which the
matter density is homogeneous and vanishes outside the sphere of radius $r=R$, that is
\be
\rho
=
\frac{3\, M_0}{4\,\pi\, R^3}\, 
\Theta(R-r)
\ ,
\label{HomDens}
\ee
where $\Theta$ is the Heaviside step function, and
\be
M_0
=
4\,\pi
\int_0^R
r^2\,\d r\,\rho(r)
\ .
\ee
Even with this static and very simple matter density, it is impossible to find analytic solutions to Eq.~\eqref{EOMV},
so we shall first solve it numerically and then proceed to find analytical approximations.
\par
For the numerical solutions, it is useful to introduce dimensionless units by considering the size $R$
of the source as the reference length, that is we define 
\be
r\equiv R\,\tilde r
\ ,
\qquad
\gn\,M_0\equiv R\,\tilde M_0
\ ,
\ee
so that Eq.~\eqref{EOMV} for $\tilde V=V(\tilde r)$ reads
\be
\left(1-4\,\tilde V\right)
\left(
\tilde \triangle \tilde V
-4\,\pi\,\tilde\rho\right)
=
2\left(\tilde V'\right)^2
\ ,
\label{tEOMV}
\ee
where $\tilde f'\equiv \d \tilde f/\d\tilde r$, the operator
$\tilde \triangle=\tilde r^{-2}\,\partial_{\tilde r}\left(\tilde r^2\,\partial_{\tilde r}\right)$ and
\be
\tilde\rho
=
R^2\,\gn\,\rho
\equiv
\frac{3\, \tilde M_0}{4\,\pi}\,\Theta(1-\tilde r)
\ .
\ee
\subsection{Newtonian solution}
As a reference, we shall first consider the Newtonian potential for this homogeneous sphere,
and then analyse the different regions of space separately for the complete Eq.~\eqref{EOMV}.
\par
Eq.~\eqref{EOMn} with the density~\eqref{HomDens} in dimensionless units reads
\be
\tilde \triangle \tilde V
=
3\,\tilde M_0\,\Theta(1-\tilde r)
\ .
\label{tEOMVn}
\ee
Its complete solution can be easily obtained by matching the (asymptotically regular) vacuum solution for $\tilde r>1$,
\be
\tilde V_{\rm N}^+
=
-\frac{\tilde M}{\tilde r}
\ ,
\ee
where $\gn\,M\equiv R\,\tilde M$, with the solution (regular in the centre) for $0\le \tilde r< 1$,
\be
\tilde V_{\rm N}^-
=
\frac{\tilde M_0}{2}\left(\tilde r^2 - 3\,C\right)
\ ,
\label{Vn-}
\ee
where $C$ is a constant to be determined.
In particular, requiring that both $\tilde V_{\rm N}$ and its derivative $\tilde V_{\rm N}'$ be continuous
across $\tilde r=1$ fixes the integration constant $C=1$ and $\tilde  M=\tilde M_0$, that is
\be
\tilde V_{\rm N}
=
\left\{
\begin{array}{lrl}
\strut\displaystyle\frac{\tilde M_0}{2}\left(\tilde r^2 - 3\right)
&
{\rm for}
&
0\le \tilde r<1 
\\
\\
-\strut\displaystyle\frac{\tilde M_0}{\tilde r}
&
{\rm for}
&
\tilde r>1
\ .
\end{array}
\right.
\label{Vn}
\ee
\subsection{Fully numerical solutions}
\label{ss:fullnum}
The above Newtonian solution $\tilde V_{\rm N}$ is obtained from the boundary conditions
\be
\left\{
\begin{array}{l}
\tilde V(\tilde r) \underset{\tilde r\to \infty}{\to} 0
\\
\\
\tilde V'(0)=0
\ ,
\end{array}
\right.
\ee
which fix all of the integration constants and only leave a dependence on $\tilde M_0=\tilde M$. 
The same boundary conditions are then required for the solution $\tilde V_{\rm c}$ to Eq.~\eqref{tEOMV},
which we therefore expect will be uniquely determined by $\tilde M_0$.
\par
\begin{figure}[t]
\centering
\includegraphics[width=8cm]{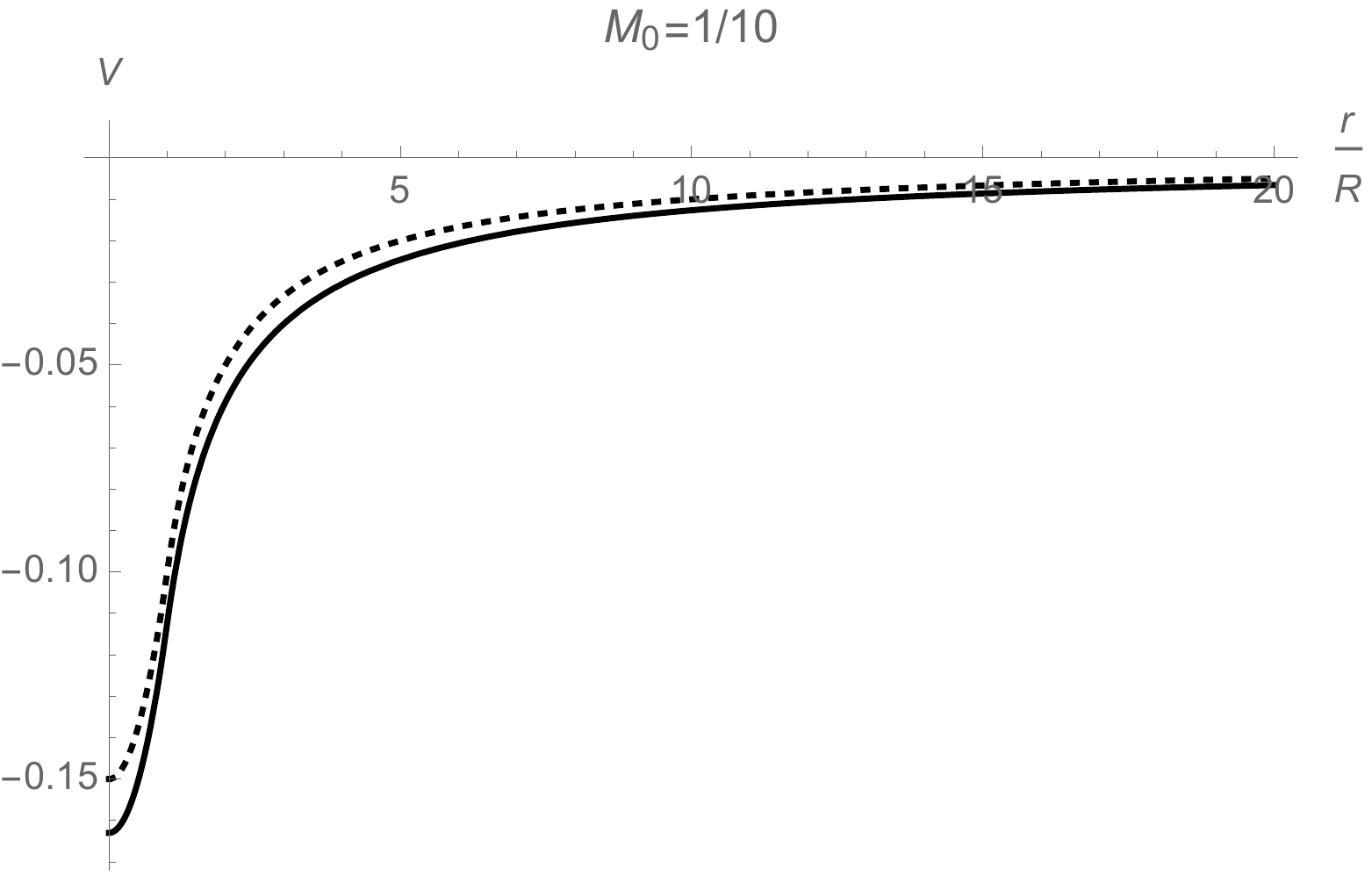}
$\ $
\includegraphics[width=8cm]{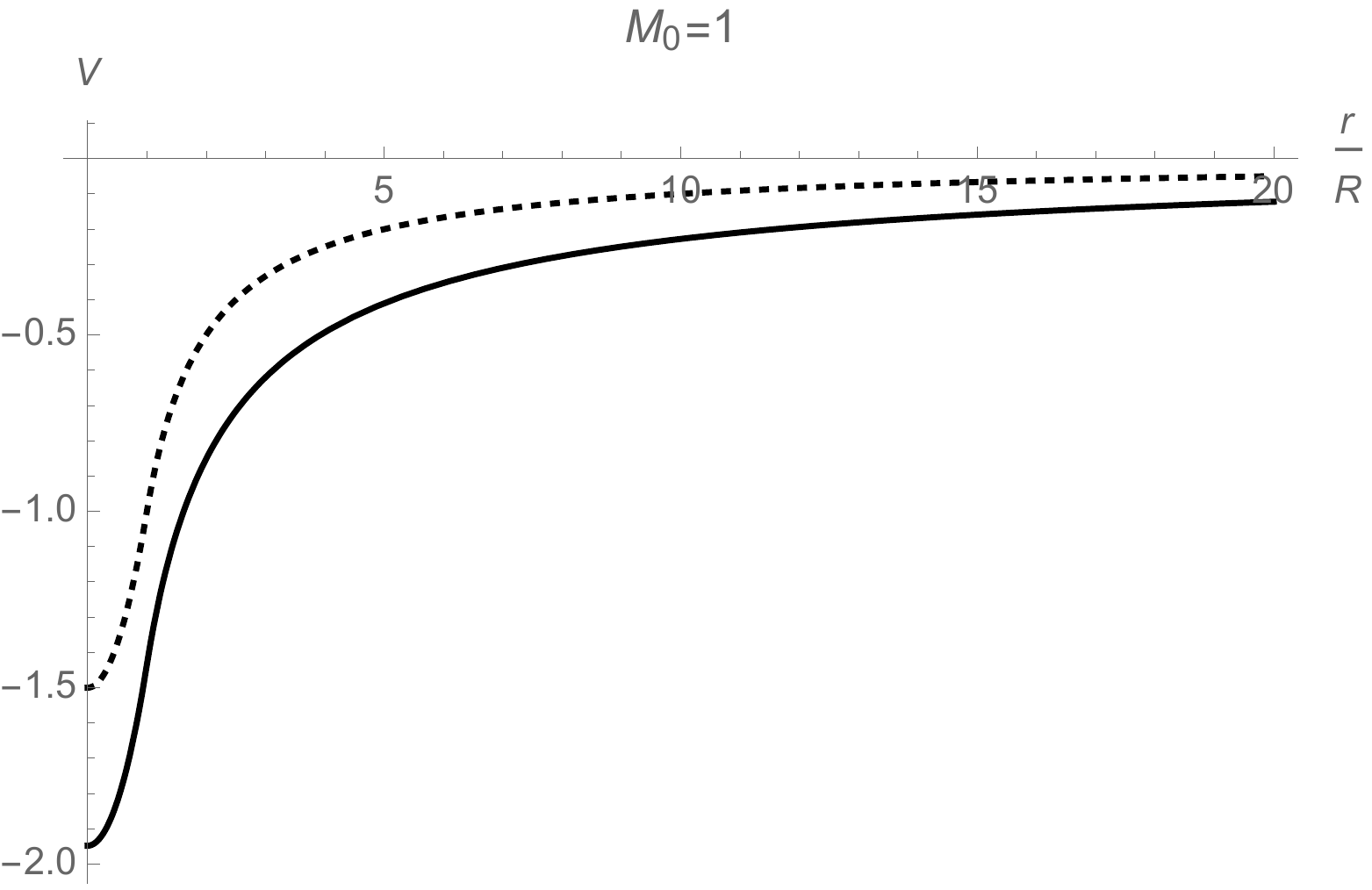}
\\
$ $
\\
\includegraphics[width=8cm]{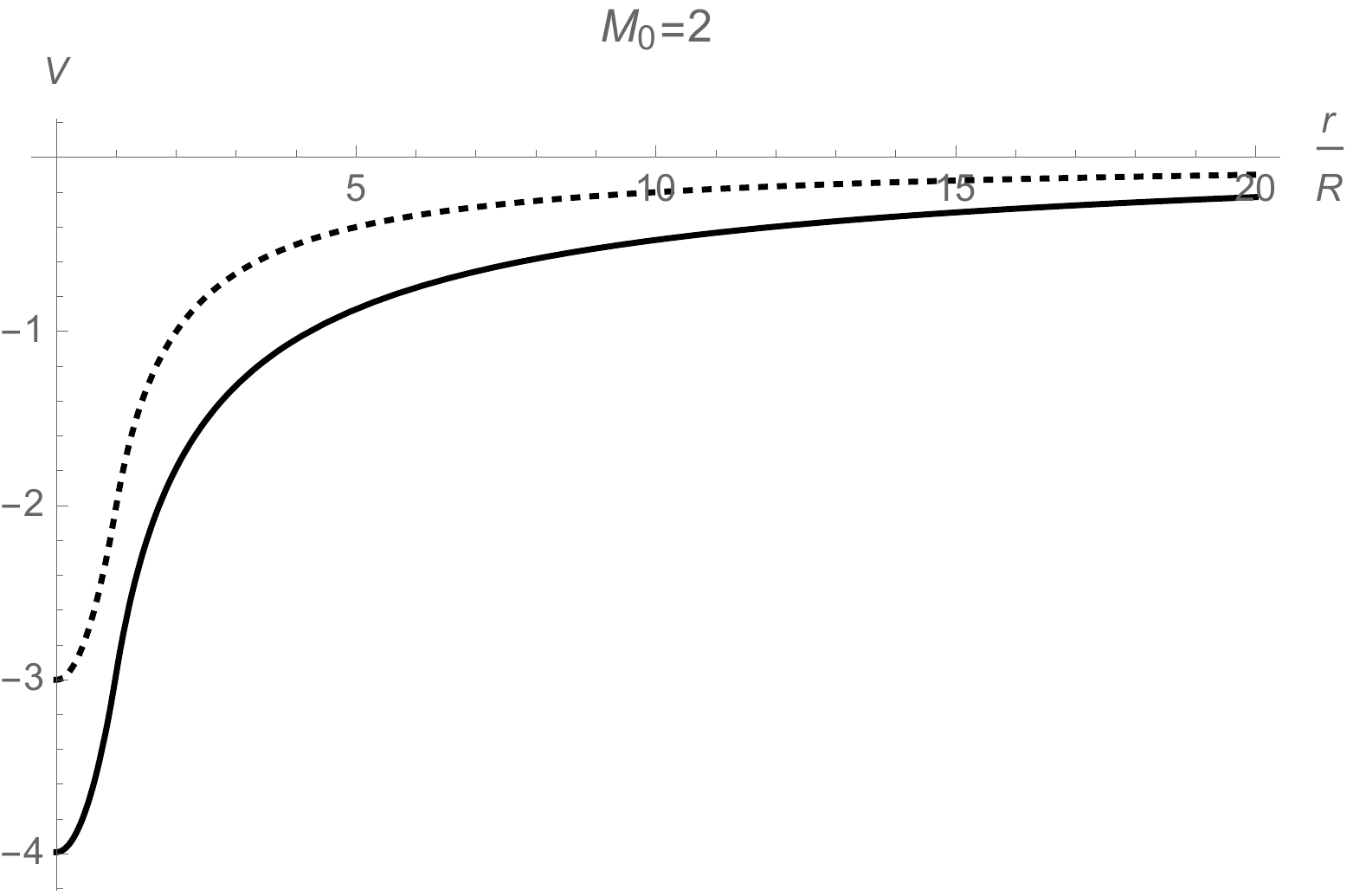}
$\ $
\includegraphics[width=8cm]{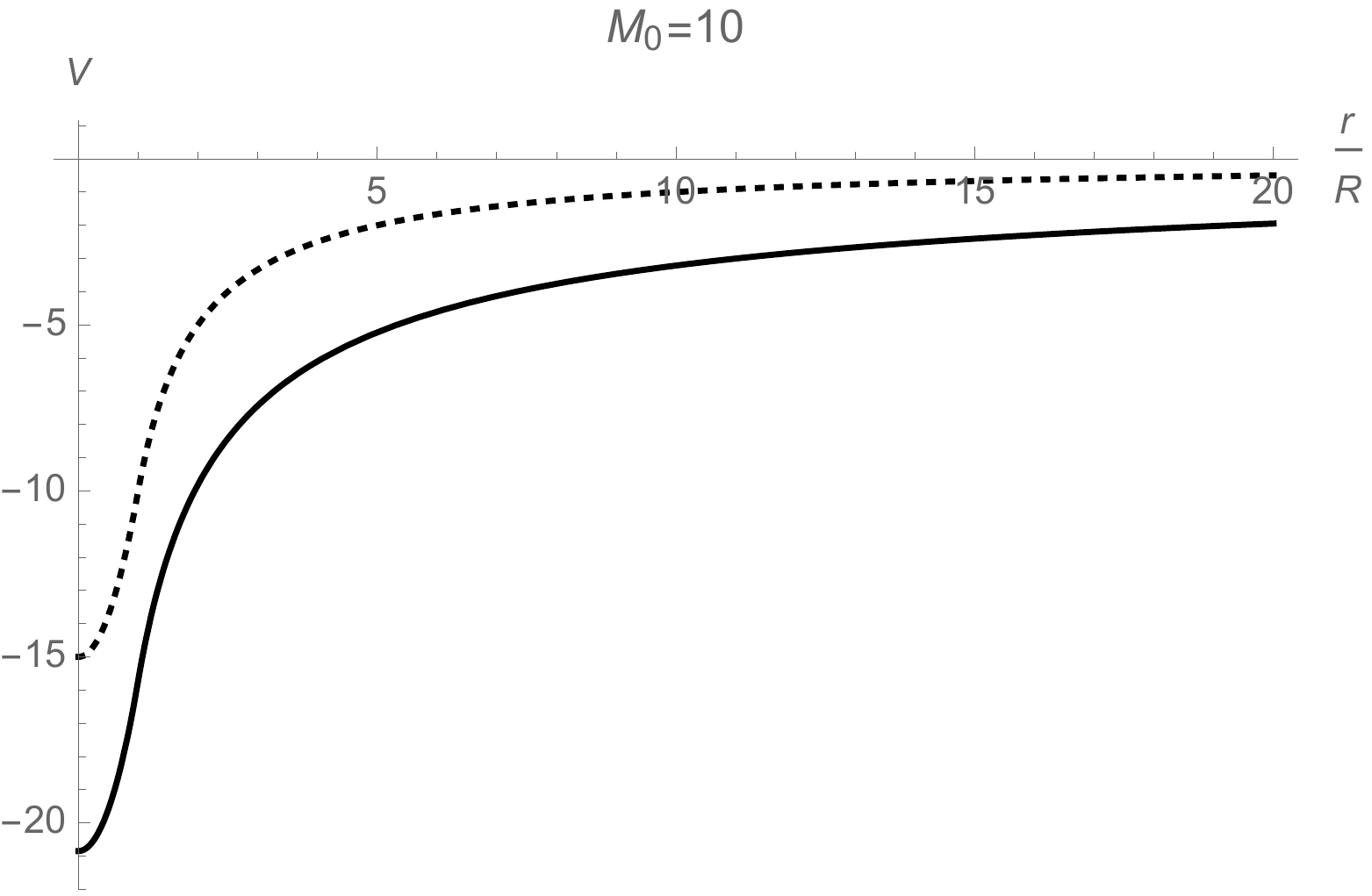}
\caption{Numerical solution $\tilde V_{\rm c}$ (solid line) {\em vs\/} Newtonian potential $\tilde V_{\rm N}$ (dashed line)
for $\tilde M_0=1/10$ (top left), $\tilde M_0=1$ (top right), $\tilde M_0=2$ (bottom left) and $\tilde M_0=10$
(bottom right).}
\label{VnVcNum}
\end{figure}

Some numerical solutions $\tilde V_{\rm c}$ computed for given values of $\tilde M_0$ are shown in
Fig.~\ref{VnVcNum}, where they are also compared with the corresponding Newtonian potentials~\eqref{Vn}.
The main features of the numerical solutions $\tilde V_{\rm c}$ are that they systematically lie below their
Newtonian counterparts $\tilde V_{\rm N}$, but their shapes are qualitatively very close.
\subsection{Analytical approximations}
The features found numerically will now guide us to find analytical approximations for the interior of the
homogeneous source.
First of all, outside the source $\rho=0$, and the exact solution~\eqref{sol0} in dimensionless units reads
\be
\tilde V_{\rm c}
=
\frac{1}{4}
\left[
1-\left(1+\frac{6\,\tilde M}{\tilde r}\right)^{2/3}
\right]
\ ,
\label{tsol0}
\ee
In the following, it will be also useful to consider the value of this potential at $\tilde r=1$,
\be
\tilde V_{\rm c}^+
=
\frac{1}{4}
\left[
1-\left(1+{6\,\tilde M}\right)^{2/3}
\right]
\ ,
\ee
and of its derivative,
\be
\tilde V_{\rm c}'^+
=
{\tilde M}
\left(1+{6\,\tilde M}\right)^{-1/3}
=
\tilde V_{\rm N}'^+
\left(1+{6\,\tilde M}\right)^{-1/3}
\ ,
\ee
where $\tilde V_{\rm N}$ is the Newtonian potential, and the term in brackets contains
the corrections to the Newtonian force at the surface of the sphere.
\subsubsection{Inside the homogeneous source} 
\label{Inside}
Let us next consider the interior of the matter source, that is $0\le \tilde r< 1$.
An exact solution for the homogeneous density can be found which is however positive 
everywhere and cannot be used for describing a ball immersed in an outer vacuum
(see Appendix~\ref{A:homo}).
\par
Since it is hard to find exact analytical solutions where the density $\rho>0$, we first
write 
\be
\tilde V_{\rm c}
=
\tilde V_{\rm N}
+
\tilde W
\ ,
\label{Va}
\ee
and assume $|\tilde W|\ll |\tilde V_{\rm N}|$ for $\tilde r\simeq 0$.
Upon replacing into Eq.~\eqref{tEOMV}, we obtain 
\be 
\tilde \triangle \tilde W
=
\frac{2\left(\tilde V_{\rm N}'+\tilde W'\right)^2}{1-4\,\tilde V_{\rm N}-4\,\tilde W}
\simeq
\frac{2\left(\tilde V_{\rm N}'\right)^2}{1-4\,\tilde V_{\rm N}}
\ ,
\label{eqW}
\ee
where $\tilde V_{\rm N}$ is given in Eq.~\eqref{Vn-}, with $C$ arbitrary.
The solution around $\tilde r=0$ can be written as
\be
\tilde W
\simeq
\frac{\tilde M_0^2\,\tilde r^4}{10\left(1+6\,C\,\tilde M_0\right)}
\left[1
+
\frac{20\,\tilde M_0\,\tilde r^2}{21\left(1+6\,C\,\tilde M_0\right)}
\right]
\equiv
\tilde W_4
+
\tilde W_6
\ ,
\label{W}
\ee
so that
\be
\left|
\frac{\tilde W_4}{\tilde V_{\rm N}}
\right|
\le
\frac{\tilde M_0}{15\,C\left(1+6\,C\,\tilde M_0\right)}
\ ,
\ee
where we assumed $C> 1$, since the numerical solutions suggest that $|\tilde V_{\rm c}(0)|>|\tilde V_{\rm N}(0)|$. 
We then find our approximation should be rather accurate for any values of $\tilde M_0$ as long as $C\ge1$.
\par
\begin{figure}[t]
\centering
\includegraphics[width=8cm]{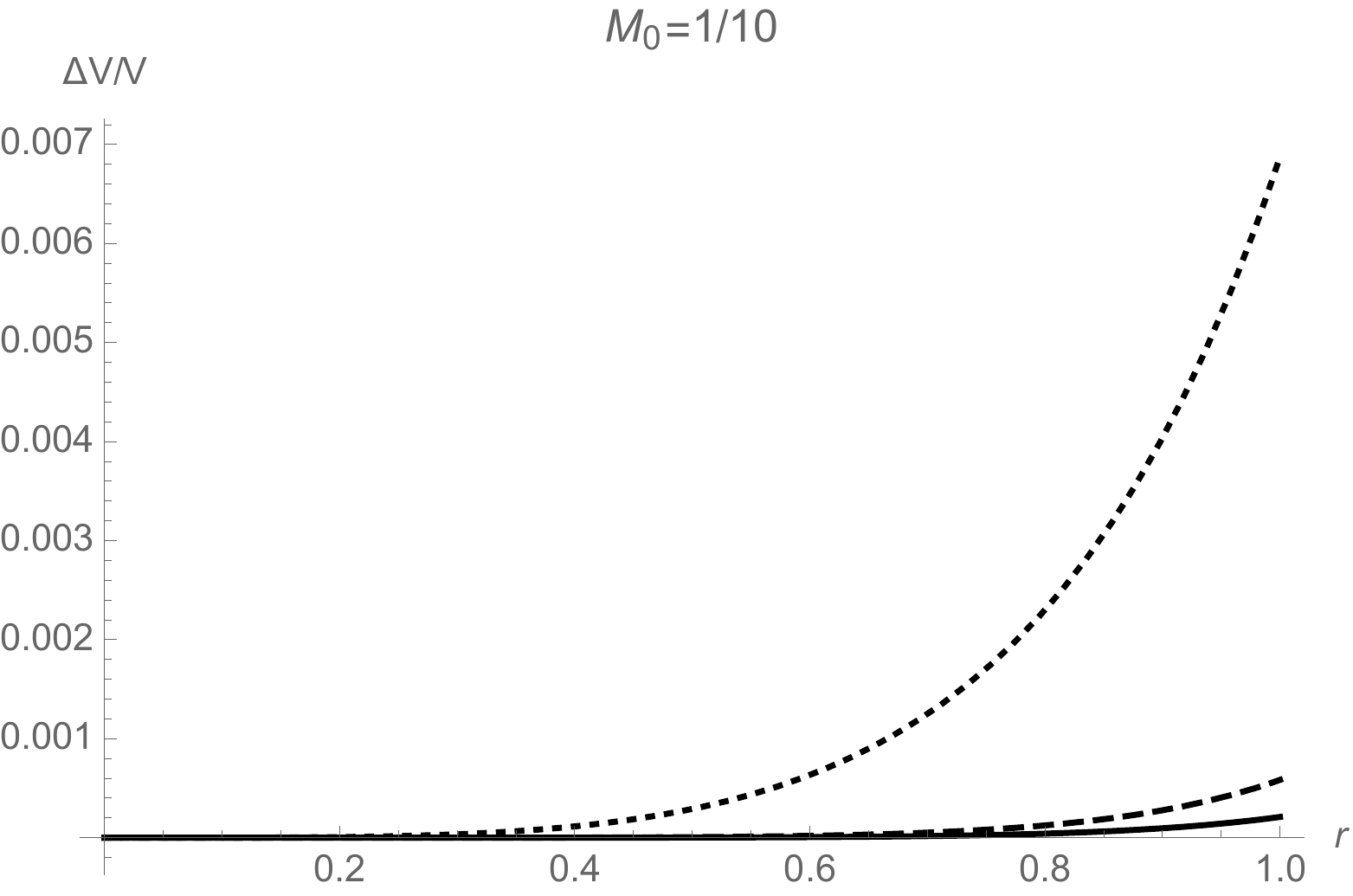}
$\ $
\includegraphics[width=8cm]{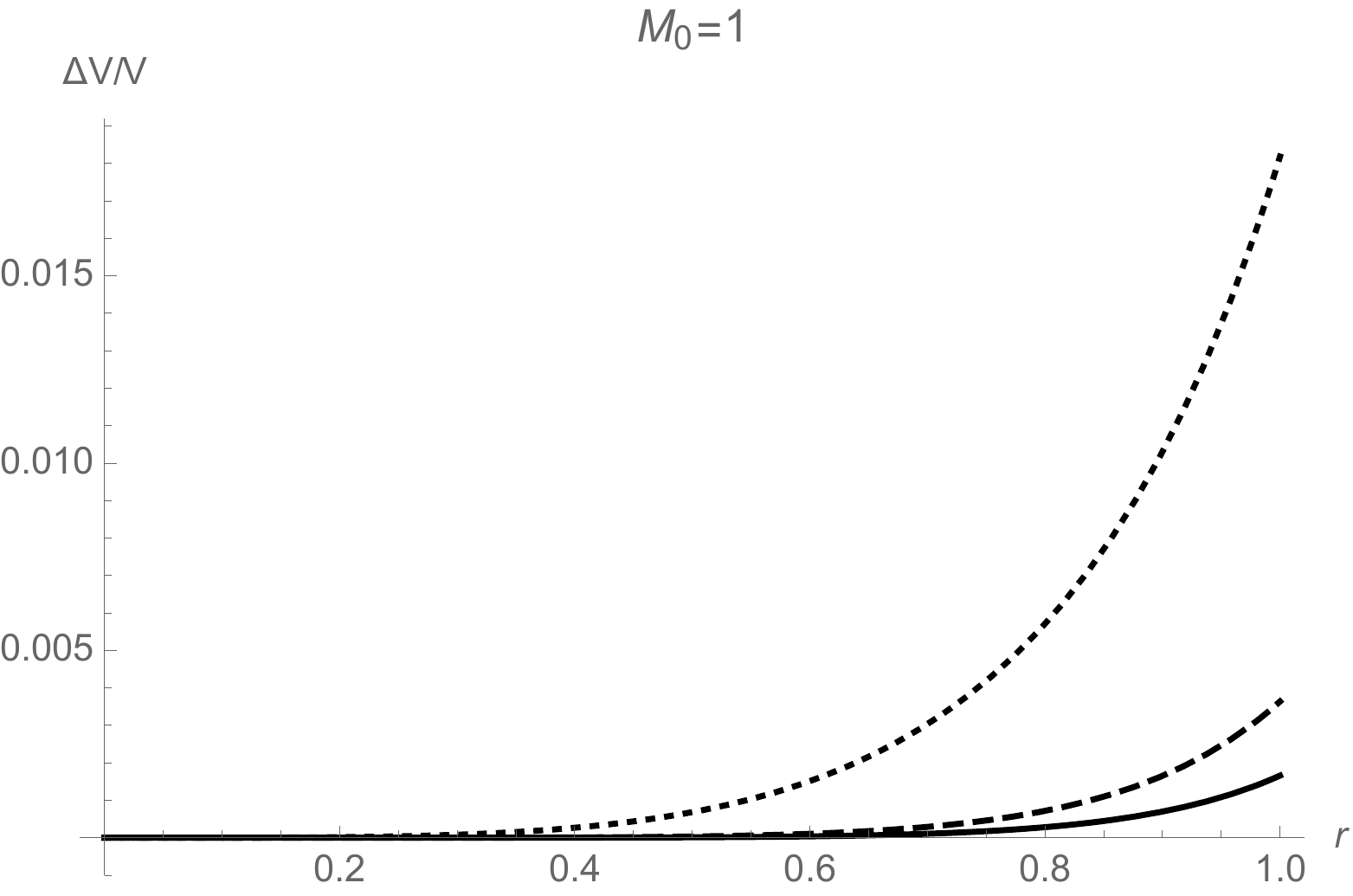}
\\ 
$\ $
\\
\includegraphics[width=8cm]{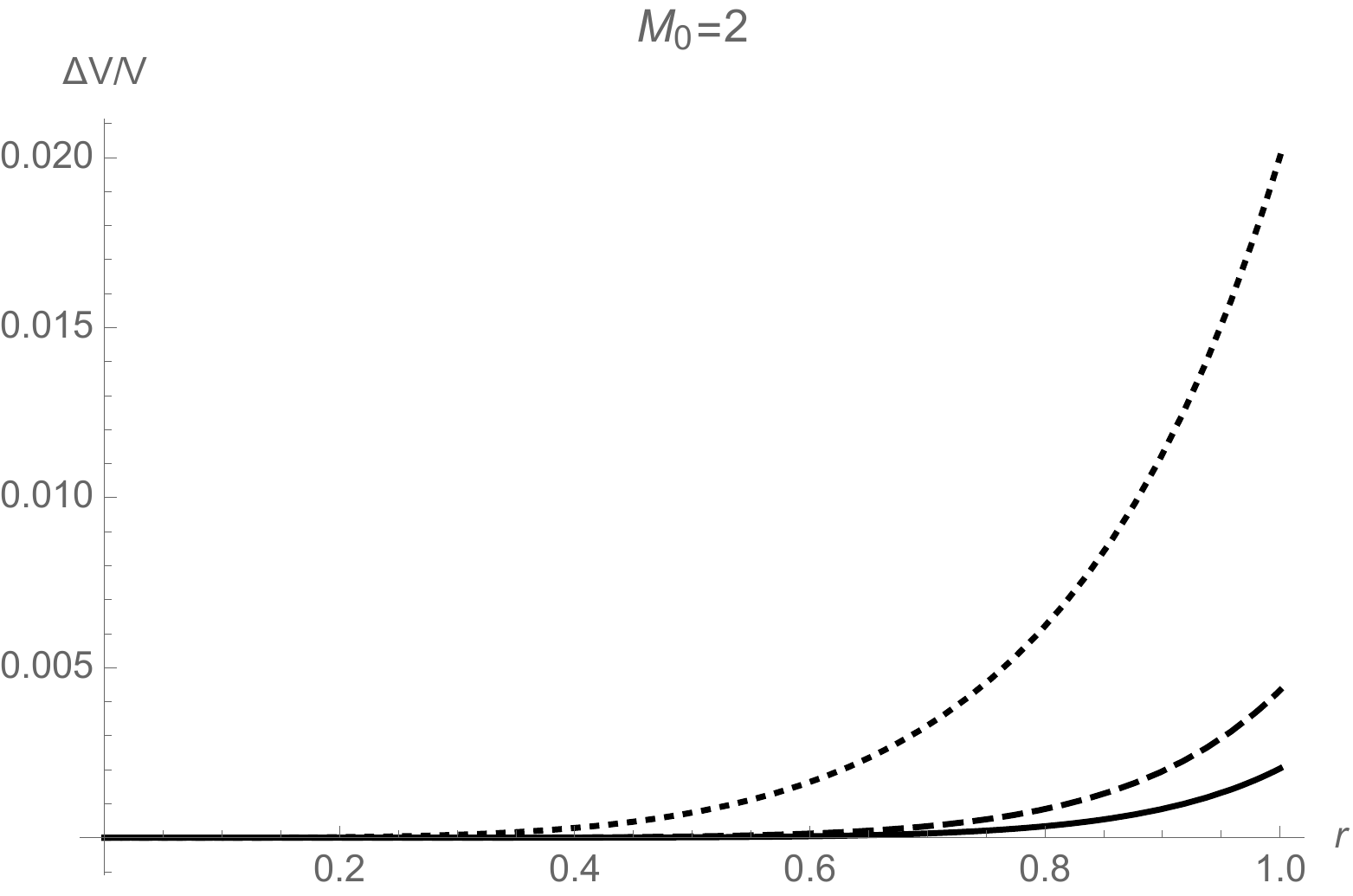}
$\ $
\includegraphics[width=8cm]{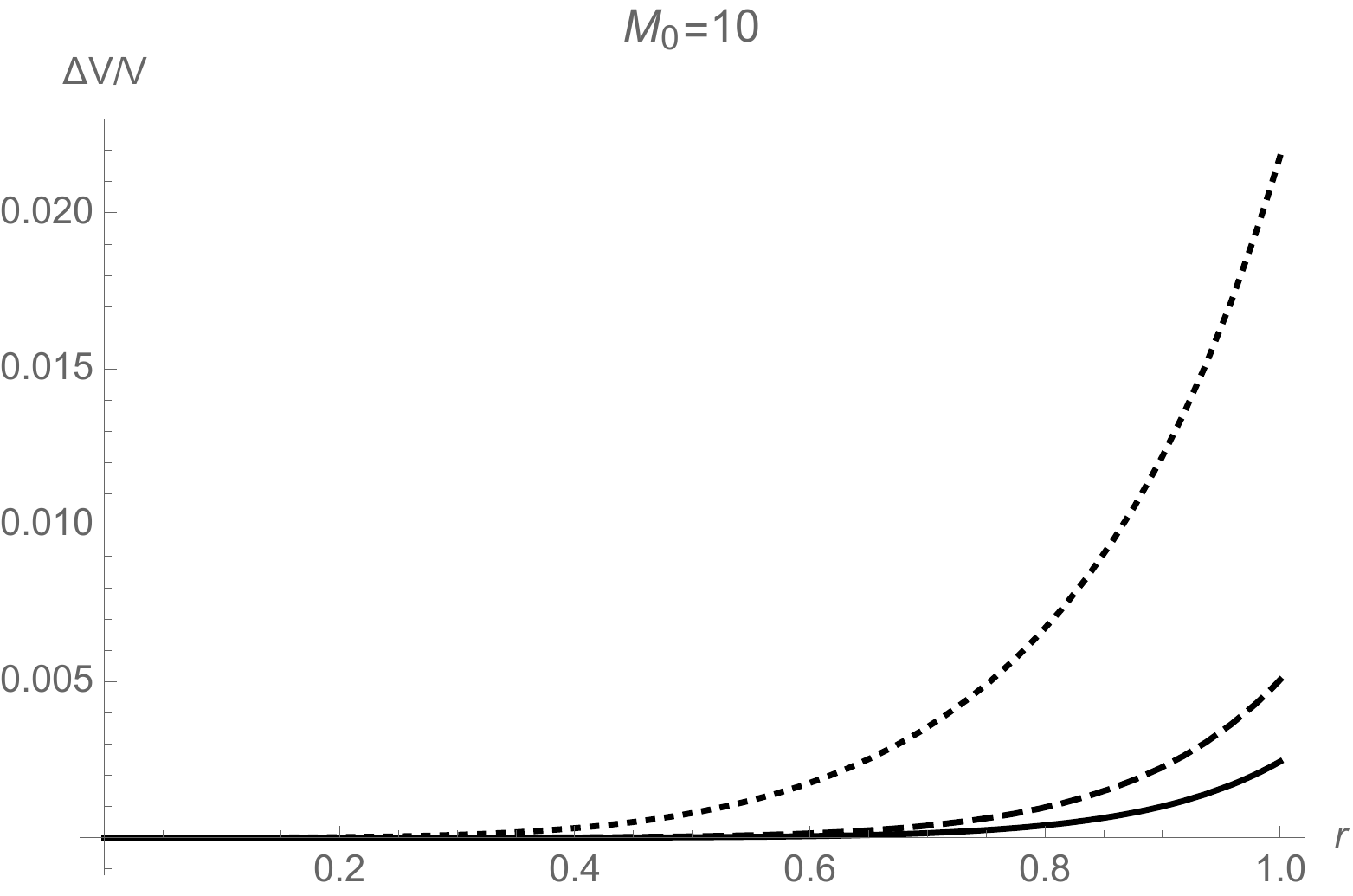} 
\caption{Relative error $(\tilde V_{\rm c}-\tilde V_{\rm n})/\tilde V_{\rm n}$ with respect to the numerical solution
$\tilde V_{\rm n}$ for $\tilde V_{\rm c}=\tilde V_{\rm N}+\tilde W_4+\tilde W_6$ (solid line) {\em vs\/}
$\tilde V_{\rm c}=\tilde V_{\rm N}+\tilde W_4$ (dashed line) {\em vs\/} Newtonian case $\tilde V_{\rm c}=\tilde V_{\rm N}$
(dotted line) for $C=1$ and $\tilde M_0=1/10$ (top left), $\tilde M_0=1$ (top right), $\tilde M_0=2$ (bottom left)
and $\tilde M_0=10$ (bottom right).
\label{ErrVc11}}
\end{figure}

The accuracy of the above analytical results is compared with the numerical solutions $\tilde V_{\rm n}$
in Fig.~\ref{ErrVc11}, which again shows that the extra source in the r.h.s.~of Eq.~\eqref{EOMV} acts as a
regulator, and whose effects are significantly dumped where $\rho> 0$.
It is important to remark that the above expansion about the Newtonian potential $\tilde V_{\rm N}$
holds for any value of $\tilde V_{\rm N}(0)$, which is determined by the constant $C$ in Eq.~\eqref{Vn-}.
Since the latter is arbitrary, and can only be fixed by matching with the outer Newtonian potential, 
the above results show that we can in fact have a solution $\tilde V_{\rm c}$ with $\tilde V'_{\rm c}(0)=0$
and $\tilde V_{\rm c}(0)$ such that it will match the outer solution~\eqref{sol0} at $\tilde r=1$. 
\subsubsection{Matching at the surface}
\begin{figure}[t]
\centering
\includegraphics[width=8cm]{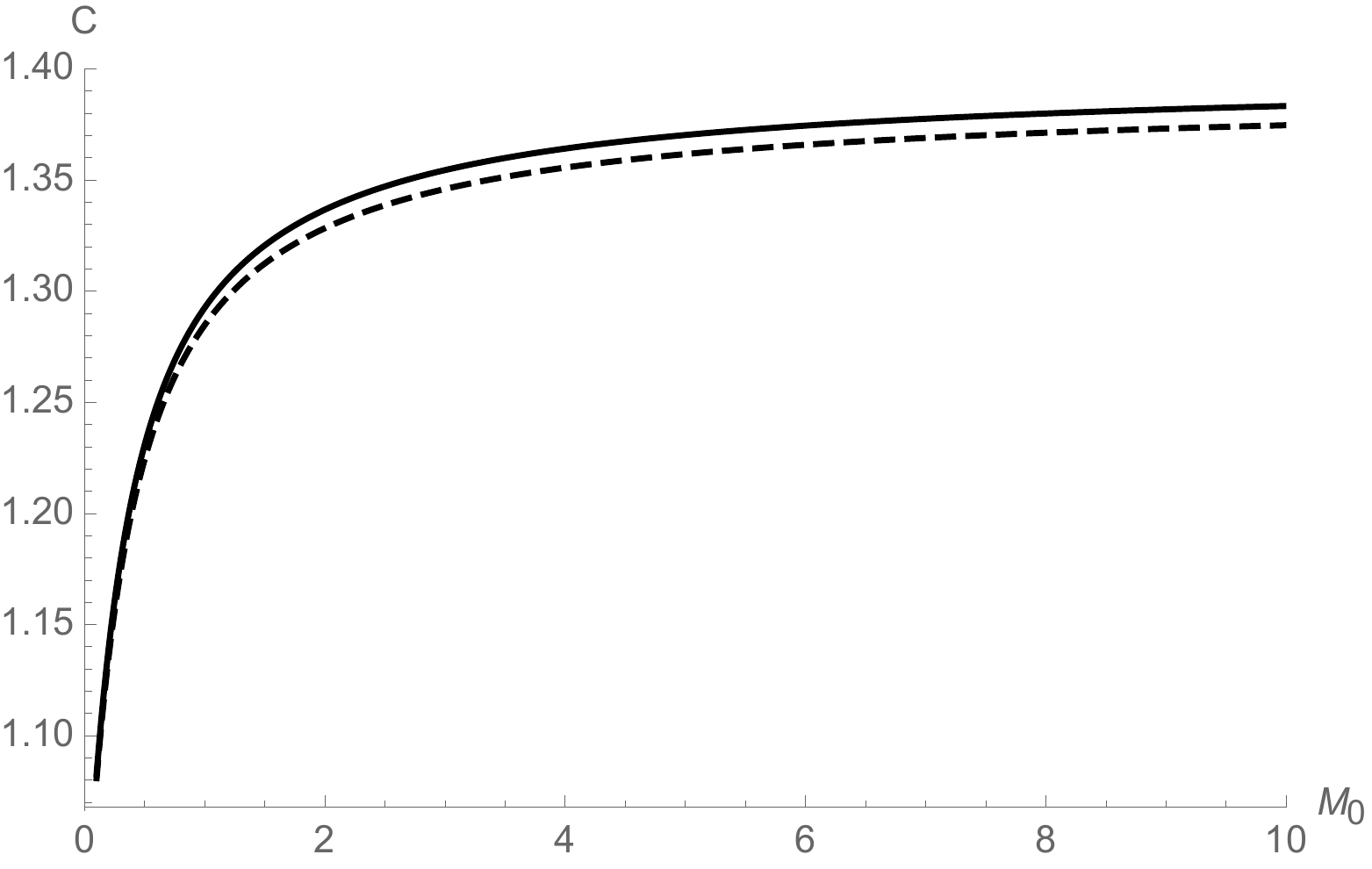}
$\ $
\includegraphics[width=8cm]{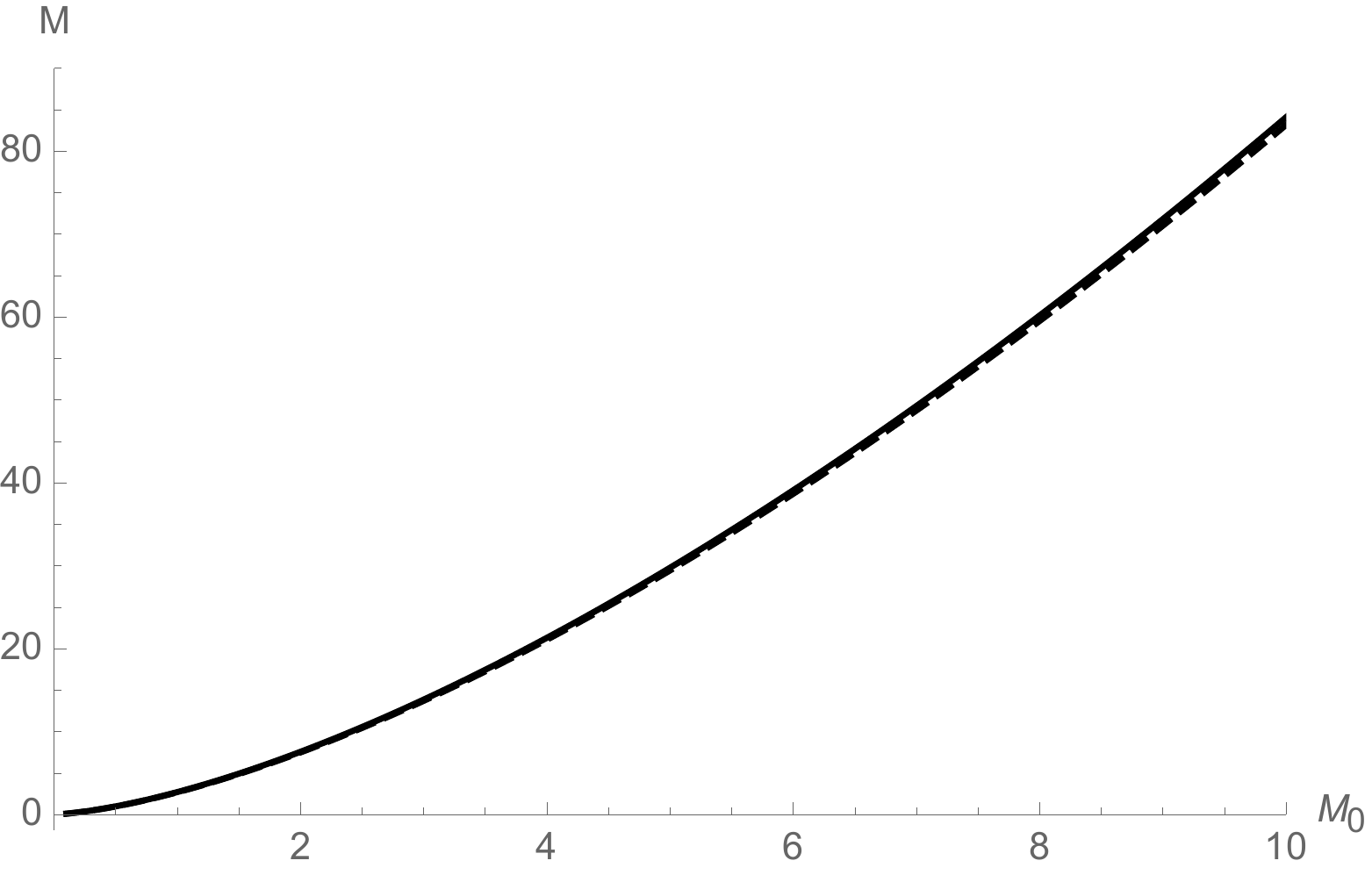}
\caption{Ratio $C=\tilde V_{\rm c}/\tilde V_{\rm N}$ (left panel) and mass $\tilde M$ (right panel) computed
numerically using $\tilde V_{\rm c}=\tilde V_{\rm N}+\tilde W_4+\tilde W_6$ (solid lines) and
using $\tilde V_{\rm c}=\tilde V_{\rm N}+\tilde W_4$ (dashed lines).
\label{CMvM0}}
\end{figure}

In order to determine the potential for all values of $\tilde r>0$, we can start from the approximate
solution~\eqref{Va} for the interior of the source and match it with the exact outer solution~\eqref{tsol0},
at $\tilde r=1$ (corresponding to $r=R$). 
In particular, with $\tilde V_{\rm N}$ in Eq.~\eqref{Vn-} and $\tilde W$ in Eq.~\eqref{W},
continuity of the potential and of its derivative at $\tilde r=1$  (see Appendix~\ref{A:junction})
results in the two conditions 
\be
\left\{
\begin{array}{l}
\tilde V_{\rm N}^-
+\tilde W^-
=
\tilde V_{\rm c}^+
\\
\\
{\tilde V}'^-_{\rm N}+{\tilde W}'^-
=
\tilde V_{\rm c}'^+
\ ,
\end{array}
\right.
\label{cont}
\ee
for the three parameters $\tilde M_0$, $\tilde M$ and $C$.
We can thus determine $C$ and $\tilde M$ in terms of $\tilde M_0$.
The numerical solutions for $C$ and $\tilde M$ in the range $1/10\le \tilde M_0\le 10$ are shown in 
Fig.~\ref{CMvM0}.
\par
Since neglecting $\tilde W_6$ does not introduce a large error, we can estimate $C$ and $\tilde M$
analytically just using $\tilde W\simeq \tilde W_4$, so that continuity of the potential at $\tilde r=1$ reads
\be
2\,{\tilde M_0}\left(1 - 3\,C\right)
+
\frac{2\,\tilde M_0^2}{5\left(1+6\,C\,\tilde M_0\right)}
\simeq
1-\left(1+{6\,\tilde M}\right)^{2/3}
\ ,
\label{con1}
\ee
whereas continuity of the derivative of the potential at $\tilde r=1$ requires
\be
\tilde M_0
+
\frac{2\,\tilde M_0^2}{5\left(1+6\,C\,\tilde M_0\right)}
\simeq
\frac{\tilde M}{\left(1+{6\,\tilde M}\right)^{1/3}}
\ .
\label{con2}
\ee
For $\tilde M_0\ll 1$, Eq.~\eqref{con2} yields the same result of Ref.~\cite{Casadio:2017cdv},
to wit
\be
\tilde M
\simeq
\tilde M_0
\left(1+\frac{12}{5}\,\tilde M_0\right)
\ ,
\label{MvM0wf}
\ee
and from Eq.~\eqref{con1} one then finds
\be
C
\simeq
1+\tilde M_0
\ ,
\label{CvM0wf}
\ee
which reproduce the Newtonian solution $C\simeq 1$ and $\tilde M\simeq \tilde M_0$ at
lowest order, and are in agreement with the numerical solutions (see, e.g.~the top left panel
of Fig.~\ref{VnVcNum}).
In the opposite limit $\tilde M_0\gg 1$, Eq.~\eqref{con2} yields
\be
\tilde M
\simeq
\sqrt{6}\,\tilde M_0^{3/2}
\ ,
\label{MvM0sf}
\ee
and Eq.~\eqref{con1} gives the asymptotic value
\be
C
=
\frac{\tilde V_{\rm c}(0)}{\tilde V_{\rm N}(0)}
\simeq
1.34
\ .
\ee
From the left panel of Fig.~\ref{CMvM0}, we see that the above estimate of $C$ is just a bit short of
the value 
\be
C\simeq 1.4
\label{Csf}
\ee
obtained from solving Eqs.~\eqref{tEOMV} numerically for $\tilde M_0\simeq 1000$.
This behaviour of $C$ and $\tilde M$ is also in agreement with the fully numerical solutions described in
Section~\ref{ss:fullnum}.
\par
It is now important to recall that $\tilde M_0=\gn\,M_0/R\gg 1$ corresponds to a source with very large
density, asymptotically approaching a Dirac delta function.
On the opposite, $\tilde M_0=\gn\,M_0/R\ll 1$ represents a source with small density for which one
expects the weak field approximation holds.
In the next Section, we will specifically look at this case.
\subsection{Weak field regime}
\label{S:weak}
The above picture simplifies significantly in the weak field regime, which was already studied in
Ref.~\cite{Casadio:2017cdv}.
In fact Eq.~\eqref{tEOMV} can be approximated for $|\tilde V|\ll 1$ as 
\be
\tilde \triangle \tilde V_{\rm WF}
\simeq
4\,\pi\,\tilde \rho
+
{2 \left(\tilde V_{\rm WF}'\right)^2}
\ ,
\label{EOMVwf}
\ee
and the main feature of this equation is that one regains the freedom to shift the potential
by an arbitrary constant, like in purely Newtonian gravity.
\par
\begin{figure}[t]
\centering
\includegraphics[width=10cm]{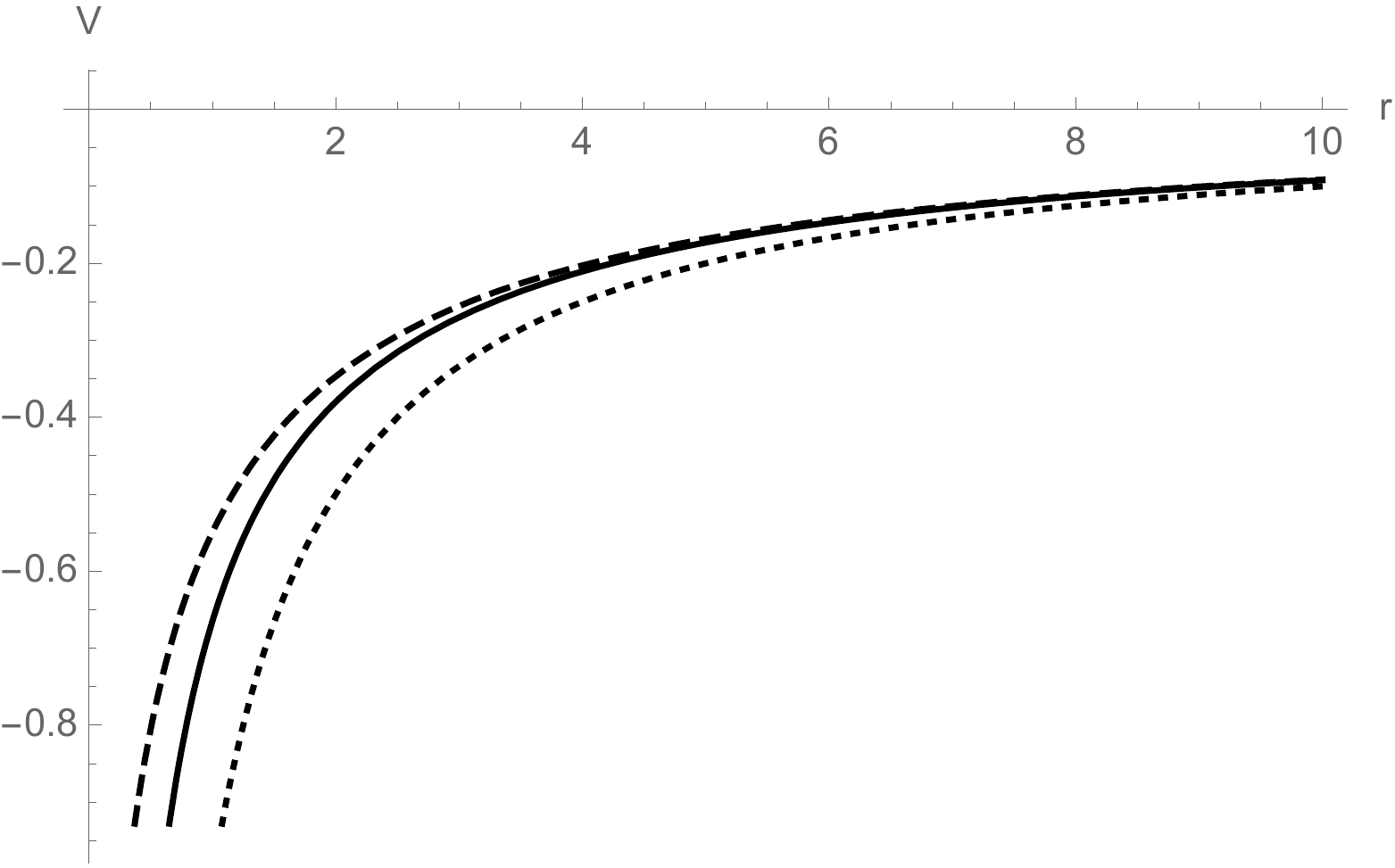}
\caption{Potential $V_{\rm WF}$ (dashed line) {\em vs\/} $V_{\rm c}$ (solid line) {\em vs\/} Newtonian potential (dotted line)
in units of $\gn\,M$.}
\label{V0new}
\end{figure}

In the vacuum, Eq.~\eqref{EOMVwf} reads
\be
\tilde \triangle \tilde V_{\rm WF}
\simeq
{2 \left(\tilde V_{\rm WF}'\right)^2}
\ ,
\label{tEOMVwf0}
\ee
and is exactly solved by
\be
\tilde V_{\rm WF}
=
-\frac{1}{2}\,\ln\left(1+\frac{2\,\tilde M}{\tilde r}\right)
\ ,
\label{Vwf0}
\ee
in which two arbitrary integration constants were fixed again by requiring the expected Newtonian behaviour
in terms of the ADM mass $\tilde M$ for large $\tilde r$.
The large $r$ expansion of the above solution is the same as Eq.~\eqref{Vlarge} up to, and including order $\gn^2$
and the correct post-Newtonian term is again recovered without any further assumptions.
In Fig.~\ref{V0new}, we plot the weak field solution~\eqref{Vwf0} and compare it to the exact non-linear
solution~\eqref{sol0} and the Newtonian potential for the same value of the ADM mass.
Clearly, the plot is only meaningful for $r\gg \gn\,M=1$ (in the units of the graph), since the condition
$|\tilde V_{\rm WF}|\ll 1$ implies that $\tilde r\gg 2\,\tilde M$.
\par
The interior of a homogeneous source is now approximately described by the equation
\be
\tilde \triangle \tilde V_{\rm WF}
\simeq
3\,\tilde M_0
+
{2 \left(\tilde V_{\rm WF}'\right)^2}
\ ,
\label{tEOMVwfIn}
\ee
which is solved exactly by
\be
\tilde V_{\rm WF}
=
A
-
\frac{1}{2}\,\ln\!\left[\frac{\sin\!\left(B+\sqrt{6\,\tilde M_0}\,\tilde r\right)}{\tilde r}\right]
\ ,
\ee
where $0\le \tilde r\le 1$ and $\tilde M_0\ll 1$ in order to preserve the weak field approximation.
Regularity in $\tilde r=0$ then fixes the integration constant $B=0$ and, upon defining
$A-{\ln(6\,\tilde M_0)}/{4}\equiv 3\,\tilde M_0\,C/2$, one finds
\be
\tilde V_{\rm WF}
=
A
-
\frac{1}{2}\,\ln\!\left[\frac{\sin\!\left(\sqrt{6\,\tilde M_0}\,r\right)}{\tilde r}\right]
\simeq
\frac{\tilde M_0}{2}
\left(\tilde r^2-3\,C\right)
+
\frac{\tilde M_0^2}{10}\,\tilde r^4
\ ,
\ee
which indeed agrees with the limit $\tilde M_0\ll 1$ of Eq.~\eqref{W}.
\par
Continuity of the derivative of the potential across $\tilde r=1$ reads
\be
\tilde M_0
\left(1+\frac{2\,\tilde M_0}{5}\right)
\simeq
\tilde M
\left(1-2\,\tilde M\right)
\ ,
\ee
and is solved by Eq.~\eqref{MvM0wf}.
Upon using this relation between $\tilde M$ and $\tilde M_0$, continuity of the potential at $\tilde r=1$
reads
\be
\frac{3\,\tilde M_0}{2}
\left(C-1\right)
\simeq
\frac{3\,\tilde M_0^2}{2}
\ ,
\ee
which is again solved by Eq.~\eqref{CvM0wf}.
This shows that for $\tilde M_0\ll 1$, one can just employ the much simpler Eq.~\eqref{EOMVwf}, like
it was done in Ref.~\cite{Casadio:2017cdv}.
\subsection{Pressure and stability}
So far we completely neglected the pressure required to have a static configuration at fixed radius $R$.
In the following, we will employ a Newtonian argument in order to estimate the necessary pressure and deduce
the corresponding potential energy contribution in a way so as to support the identification of $M$ as the total
energy of the system.
\subsubsection{Newtonian pressure}
It is well know from the Newtonian theory that the condition of hydrostatic equilibrium for the pressure
$p_{\rm N}$ reads~\cite{weinberg}
\be
p_{\rm N}'(r)
=
-V_{\rm N}'(r)\, \rho(r)
=
-\frac{\gn\, m(r)}{r^2}\, \rho(r)
\ , 
\label{pressureeq}
\ee
where $m=m(r)$ is the mass contained within a sphere of radius $r$ defined in Eq.~\eqref{m(r)}.
This equation can be easily solved for the homogeneous density~\eqref{HomDens} generating the
Newtonian potential~\eqref{Vn} inside the source.
Upon requiring that the pressure vanishes at the surface, $p_{\rm N}(R)=0$, the solution reads
\be
p_{\rm N}(r)
=
\frac{3\, \gn\, M_0^2\,(r^2-R^2)}{8\ \pi\ R^6}
\ .
\label{pN}
\ee
It is then straightforward to calculate the associated (Newtonian) potential energy as
\be
U_{\mathrm{BN}}(r)
&\!\!=\!\!&
B_{\rm N}- 4\, \pi \int_0^r\, \d \bar r\, \bar r^{2}\, p_{\rm N}(\bar r)
\nonumber
\\
&\!\!=\!\!&
B_{\rm N}
+
\frac{\gn\, M_0^2\,(3\,r^2-5\,R^2)\,r^3}{10\, R^6}
\ ,
\label{Ub}
\ee
where $B_{\rm N}$ is an arbitrary integration constant.
\par
The constant $B_{\rm N}$ can in fact depend on the parameters of our model, like $M_0$ and $R$.
In particular, the condition of stability requires that the work done by the pressure $p_{\rm N}$
cancels against the work done by the gravitational force for an infinitesimal change in the radius $R$,
that is
\be
\frac{\d U_{\rm{BN}}}{\d R}
=
-\frac{\d U_{\mathrm{N}}}{\d R}
=
-\frac{3\, \gn\, M_0^2}{5\, R^2}
\ ,
\ee
in which we used Eq.~\eqref{UN} for the Newtonian potential energy $U_{\rm N}=U_{\rm N}(R)$.
This condition yields
\be
B_{\rm N}(M_0,R)
=
\frac{4\, \gn\, M_0 ^2}{5\, R}
\ ,
\label{B0}
\ee
and
\be
U_{\mathrm{BN}}(r)
=
\frac{\gn\, M_0^2\,(3\,r^5-5\,r^3\,R^2+8\,R^5)}{10\, R^6}
\ ,
\label{Ubn}
\ee
which ensures that the total energy $E$ of the system
\be
E=M_0+U_{\rm BN}(R)+U_{\rm N}(R)
=
M_0
\ .
\ee
Moreover, it is worth recalling that $M_0=M$ at the Newtonian level, so that $E=M$.
\subsubsection{Post-Newtonian pressure}
\label{ss:pnpressure}
\begin{figure}[t]
\centering
\includegraphics[width=8cm]{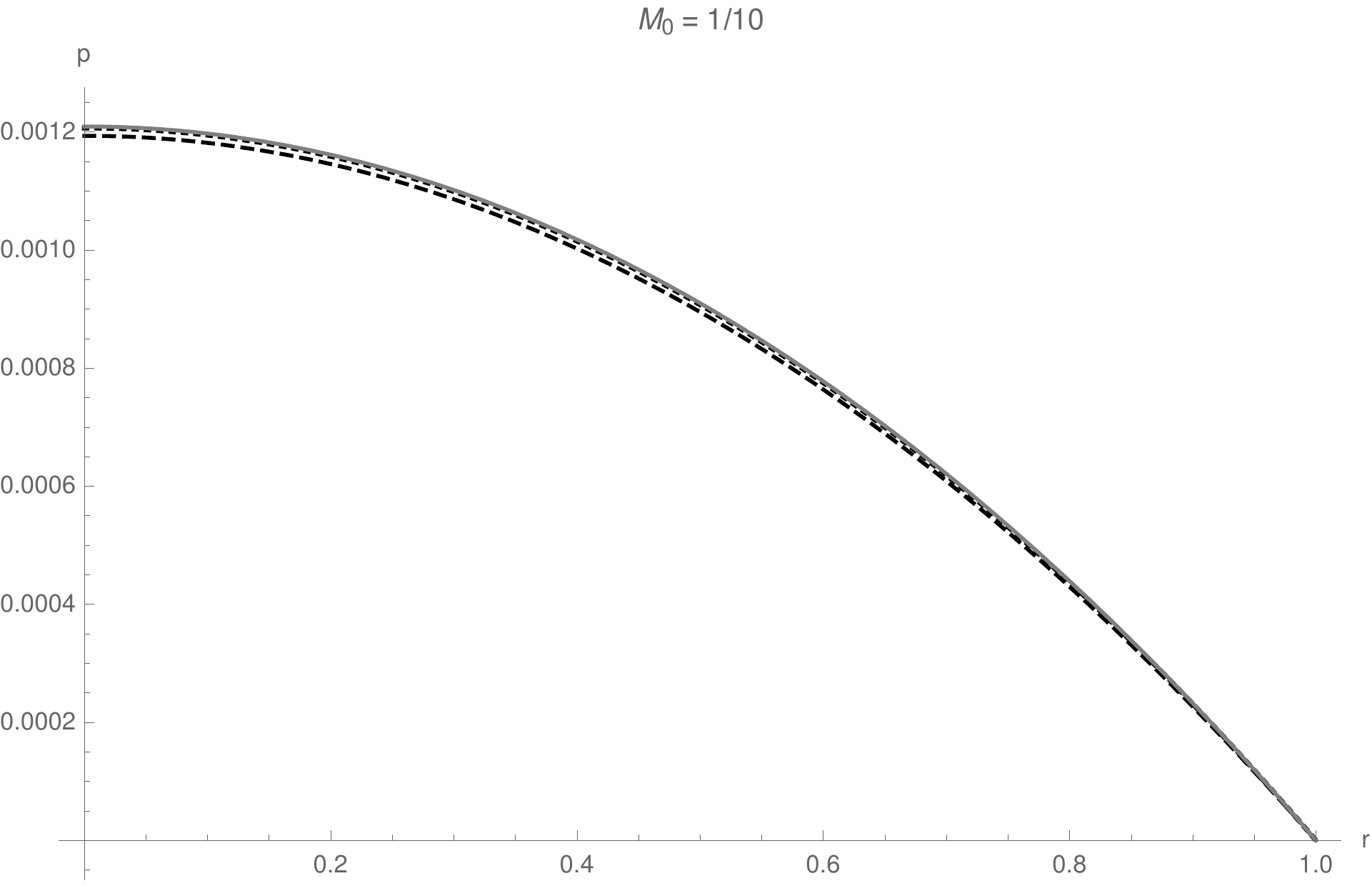}
$\ $
\includegraphics[width=8cm]{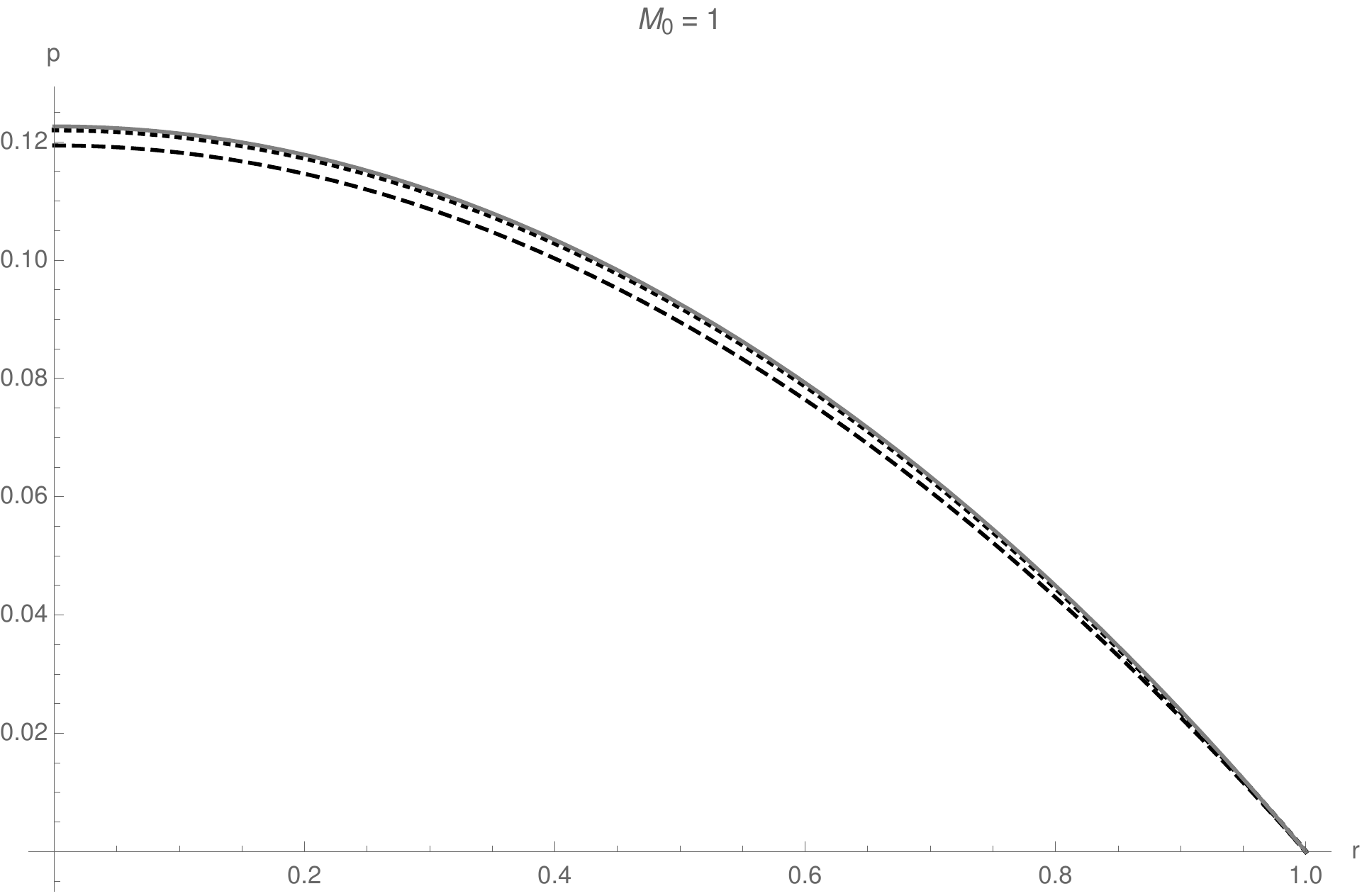}
\\
$ $
\\
\includegraphics[width=8cm]{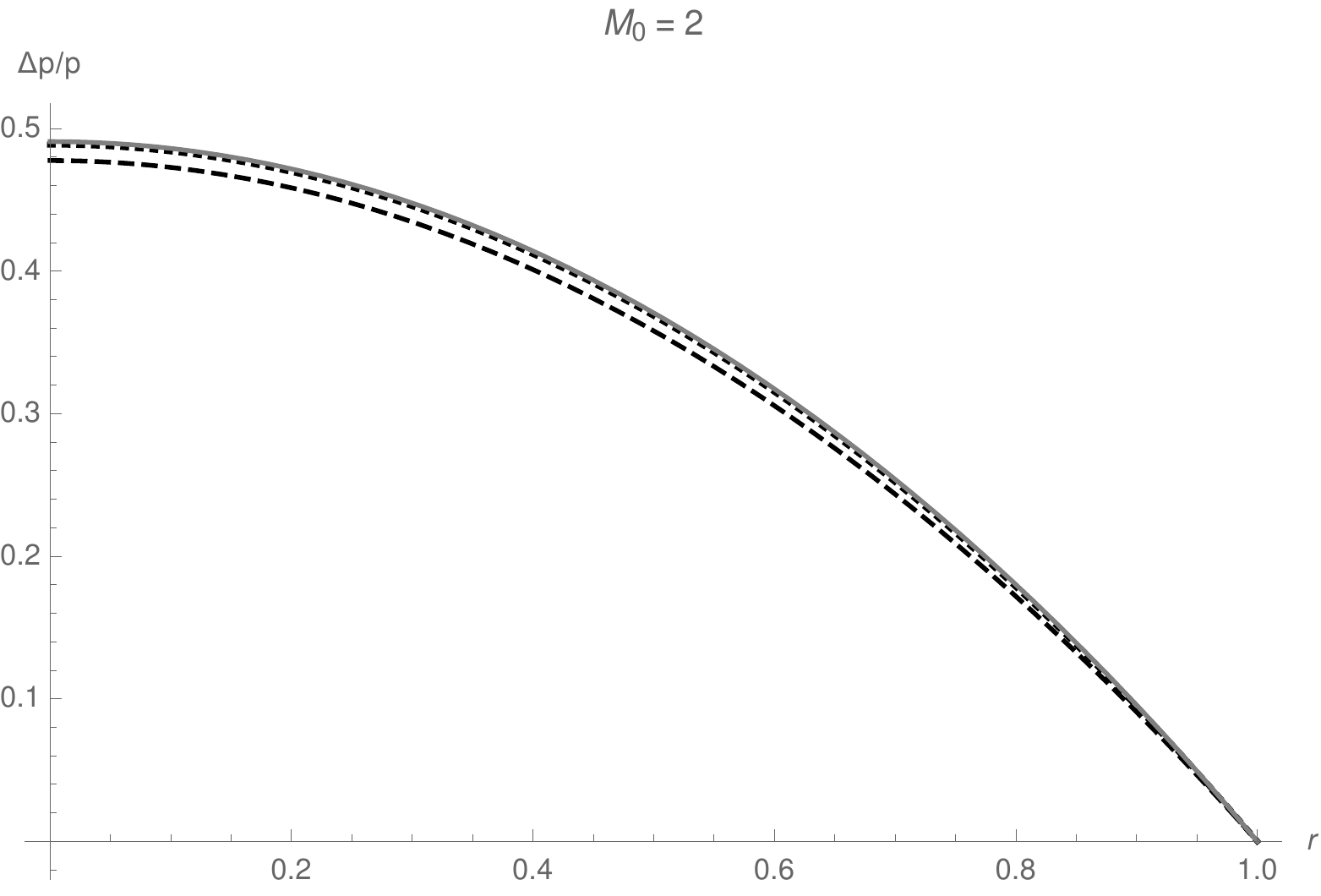}
$\ $
\includegraphics[width=8cm]{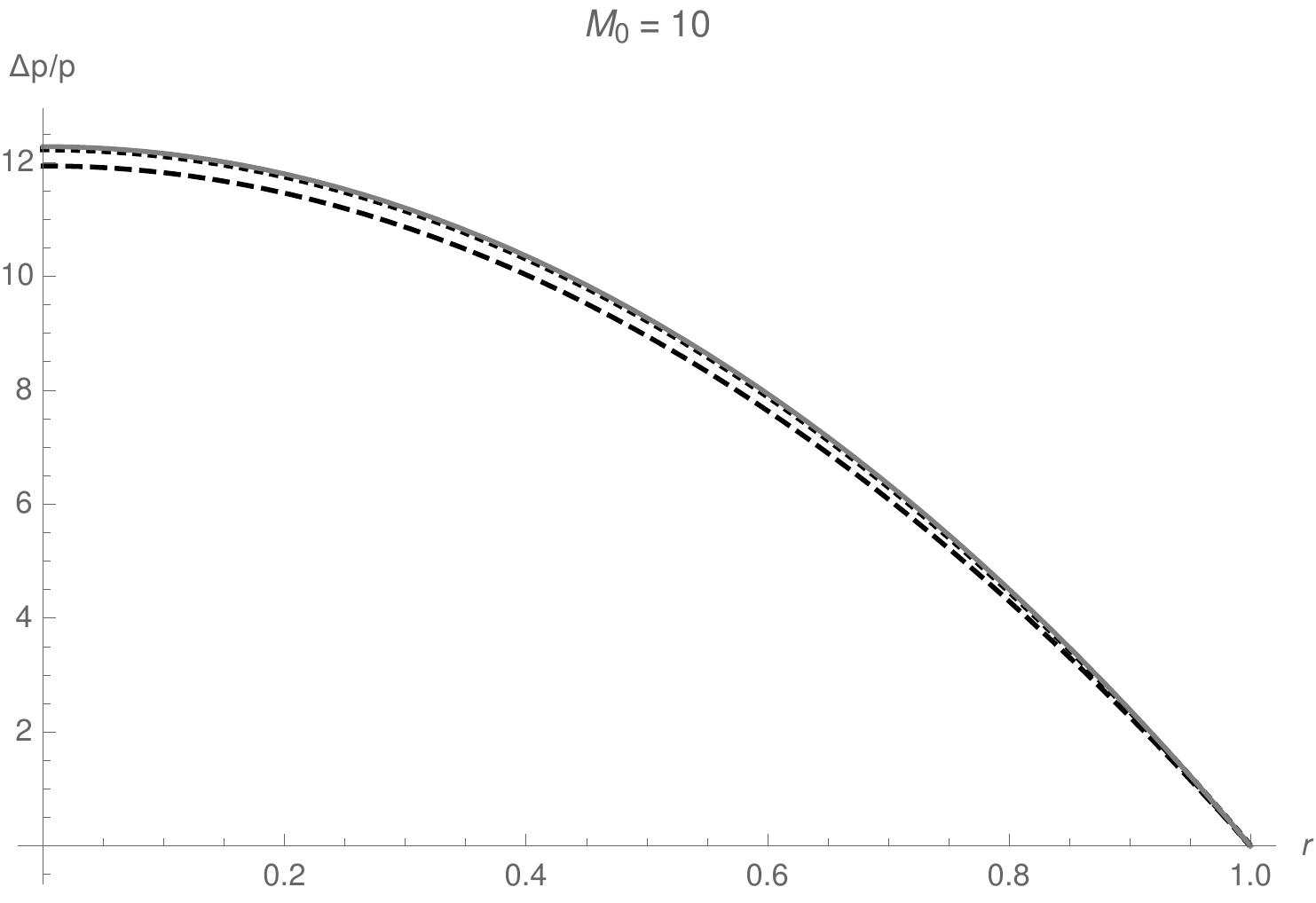}
\caption{Numerical solution (solid gray line) {\em vs\/} analytical approximation~\eqref{pp} with $C=1.34$ (dotted line)
{\em vs\/} Newtonian pressure~\eqref{pN} (dashed line) for $\tilde M_0=1/10$ (top left), $\tilde M_0=1$ (top right),
$\tilde M_0=2$ (bottom left) and $\tilde M_0=10$ (bottom right).}
\label{P}
\end{figure}

For the configurations found in Section~\ref{S:solution}, we follow the same line of reasoning and simply
replace the Newtonian potential in Eq.~\eqref{pressureeq} with the solution $V_{\rm c}$, to wit
\be
p'(r)
=
-V'_{\mathrm{c}}(r)\, \rho(r)
\ .
\label{p'c}
\ee
For the potential $V_{\rm c}$, we shall employ the analytical approximation~\eqref{Va} up to the term
$W_4$ in Eq.~\eqref{W}, for which Eq.~\eqref{p'c} can be solved to obtain
\be
p(r)
=
\frac{3\, \gn\, M_0 ^2\,(R^2-r^2) \{5\,R^3+\gn\, M_0\,[r^2 + (1+30\, C)\,R^2]\}}
{40\, \pi\, R^8\, (R+6\,C\, \gn M_0)}
\ ,
\label{pp}
\ee
which is very close to the fully numerical result, as shown in Fig.~\ref{P}.
\par
Like in the Newtonian approximation, upon integrating the above pressure on the volume of the source,
we can obtain the baryonic potential energy $U_{\mathrm{B}}$ necessary for the compact object to be
in mechanical equilibrium,
\be 
U_{\rm{B}}(r)
&\!\!=\!\!&
\frac{\gn\, M_0^2\,\{3\,\gn\,M_0\, r^7+21\,r^5\, R^2\, (R+6\, C\, \gn\, M_0)
-7\, r^3\, R^4\, [5\,R+\gn\, M_0\,(1+30\, C) ]\}}
{70\, R^8\, (R+6\, C\, \gn\, M_0)}
\nonumber
\\
&&
+
B(M_0, R)
\ ,
\label{UBr}
\ee
where $B$ is again an integration constant, which can be fixed like in the Newtonian case, so as to ensure that $M$
equals the total energy of the system,
\be
\label{EM}
E \simeq M_0 + U_{\mathrm{G}} + U_{\mathrm{B}}
=
M
\ .
\ee
\par
The cumbersome expression~\eqref{UBr} simplifies significantly if we consider a source with small compactness,
{\em i.e.}~$\gn\,M_0/R \ll 1$ and $C \simeq 1 + \gn\,M_0/R$, which yields
\be
U_{\rm{B}}(R)
\simeq
B(M_0, R)
-\frac{\gn\, M_0^2}{5\, R}
-\frac{2\, \gn^2\, M_0^3}{35\, R^2}
\ .
\ee
In this low compactness regime, $M$ is given in Eq.~\eqref{MvM0wf} and $U_{\mathrm{G}}$ by the first line of
Eq.~\eqref{Uapp}.
Upon replacing these expressions in Eq.~\eqref{EM} then yields
\be
B(M_0, R)
\simeq
\frac{16\, \gn\,M_0 ^2}{5\, R}
+\frac{113\, \gn^2\, M_0 ^3}{70\, R^2}
\ ,
\ee
and we finally end up with 
\be
U_{\rm{B}}(R)
\simeq
\frac{3\, \gn\, M_0^2}{R}
+
\frac{109\, \gn^2\, M_0^3}{70\, R^2} 
\ .
\ee
\par
As we pointed out in Section~\ref{Inside}, our description let us also consider the opposite regime,
{\em i.e.}~the case of highly compact sources with $\gn\,M_0/R \gg 1$.
The corresponding baryonic potential energy then reads
\be
U_{\rm{B}}(R)
\simeq
B(M_0, R) 
- 
\frac{\gn\,(1  + 21\, C)\, M_0 ^2}{105\, C\, R} 
\ ,
\ee
and it is again straightforward to ensure that Eq.~\eqref{EM} is satisfied,
by fixing the integration constant 
\be \nonumber
B(M_0, R)
\simeq
2.22\,\frac{\gn^2\, M_0^3}{R^2}
\ ,
\ee
where we used the second expression in~\eqref{Uapp} for the gravitational energy $U_{\mathrm{G}}$ and Eq.~\eqref{MvM0sf} for $M$.
It is finally immediate to write
\be
U_{\rm{B}}(R) 
\simeq
2.22\,\frac{\gn^2\, M_0 ^3 }{R^2} 
\ .
\ee
\par
We would like to end this section with some remarks about the ratio $\Gamma_0\equiv p(0) / \rho(0)$. 
In General Relativity, it is well known~\cite{stephani} that, for any realistic spherical configuration of matter~\footnote{Non-outward
increasing matter density and vanishing pressure at the surface.},
the pressure can be positive and non-singular in the origin only if the radius of the source is larger than
the Buchdahl limit, that is for a source of size
\be
R
> 
\frac{9}{4}\, \gn\, M_0
\equiv 
R_{\rm BL}
\ .
\label{Rbl}
\ee
This is apparent from the expression of the pressure~\eqref{ptov} obtained by solving the
Tolman-Oppen\-heimer-Volkoff (TOV) equation~\eqref{tov} for a homogeneous density.
If we instead calculate the above ratio for our case using the pressure from Eq.~\eqref{pp},
also obtained for a homogeneous density, we find
\be
\Gamma_0
=
\frac{\gn\,M_0}{2\,R}
\left[1
+\frac{\gn\,M_0}{5\left(R+6\,C\,\gn\,M_0\right)} 
\right]
\ .
\label{gamma_0}
\ee
For $R\simeq R_{\rm BL}$, and using the value of the constant $C$ from Eq.~\eqref{Csf}, this becomes 
\be
\Gamma_0(R_{\rm BL})
\simeq
0.22
\ ,
\ee
whereas $\Gamma_0\to\infty$ in the limit $R\to 0$.
This means that the divergence in the pressure that one encounters for $R=R_{\rm BL}$ in the
General Relativistic case has been pushed all the way down to $R\simeq 0$ in our model. 
\par
The above result leads us to remark that, although a static object with a uniform density
is just a simple toy model, it can give us hints for more general arguments.
For any density profile, the gravitational force is directed inwards and would collapse the mass
towards the centre unless it were balanced by an equal and opposite pressure.
For the homogeneous density we calculated this necessary pressure without discussing its origin
or physical acceptability.
In General Relativity, only for objects with radii larger than the Buchdahl limit, such as stars, 
this pressure could be provided by nuclear reactions or the Pauli exclusion principle:
below the Buchdahl limit there exists no density profile which allows for a static
configuration.
This conclusion could only change if gravity departs from the General Relativistic description and
provides itself an effective pressure directed outwards, like in the de~Sitter space-time.
In the corpuscular picture, the gravitons inside black holes form a Bose-Einstein condensate, whose 
equation of state resembles that of dark energy~\cite{DvaliGomez}, and these gravitons are therefore natural
candidates to provide an effective outward pressure.
Moreover, such a large deviation of the ratio $\Gamma_0$ from General Relativity as the one
we found above would imply that quantum effects have overcome the classical behaviour already at
(relatively) large scales $R\simeq R_{\rm BL}\sim \gn\,M_0$.
Of course, one could then argue that this is only acceptable for masses $M_0$ sufficiently small,
say around the Planck scale.
However, one could also argue that the difference between the bootstrapped Newtonian description
and General Relativity should become smaller if one included higher interaction terms in the
Lagrangian~\eqref{LagrV}, to the point that this difference might become phenomenologically negligible
for astrophysical objects.
Finally, we also recall that, unlike what happens in General Relativity, we assumed the potential energy
density responsible for the pressure does not contribute a source term for the potential, and the addition
of such a contribution to the Lagrangian~\eqref{LagrV} might again reduce the difference with respect
to General Relativity.
\section{Conclusions and outlook}
\setcounter{equation}{0}
\label{S:conc}
In the previous sections, we have derived the potential satisfying the non-linear equation~\eqref{EOMV}
generated by a spherically symmetric homogenous compact source.
We have also obtained the necessary pressure~\eqref{pp} to keep the configuration static for sources of
arbitrarily small radii, including values below the Buchdahl limit~\eqref{Rbl}.
Since this pressure turns out to be finite, unlike what one expects in General Relativity, it now seems
appropriate to come back to our original  motivation for constructing such a ``bootstrapped'' Newtonian
ball and further clarify the possible connection with the corpuscular picture of black holes that we have
mentioned in the Introduction and at the end of the last Section.
\par
First of all, it is a theorem in General Relativity that, under rather general assumptions, systems that
develop trapping surfaces will collapse into singularities~\cite{hawkingEllis}.
This is what one expects would happen to a body that shrinks below the Buchdahl limit~\eqref{Rbl}.
On the other hand, a singularity is hardly acceptable in the quantum theory, just because of the
Heisenberg uncertainty principle, and one could generically expect that the actual collapse of an astrophysical
body will necessarily deviate from the General Relativistic description at some point.
The crucial question is whether such deviations occur after the horizon has appeared, and they therefore
remain hidden forever, or the quantum effects induce departures from General Relativity outside the 
gravitational radius which can therefore be observed.
Many works have shown the existence of regular black hole solutions of modified gravitational equations
which entail no significant departures from the corresponding General Relativistic space-times outside
the (outer) horizon (for a review, see Ref.~\cite{nicolini}). 
The corpuscular picture~\cite{DvaliGomez} instead assumes that black holes are fully quantum objects
in order to give a consistent description of the Hawking evaporation and the possibility of observable
consequences should therefore not be excluded.
\par
In order to gain further insight into the above two possibilities, one should not forget about the matter that
collapses and causes the emergence of the black hole geometry.
In particular, one might wonder where this matter ends up in the corpuscular picture~\cite{Casadio:2016zpl}.
Clearly, if General Relativity remains a good theory of gravity up to extremely high energy densities,
the collapsing matter should form a tiny ball with essentially no modifications of physics below 
the Planck energy scale.
In the quantum corpuscular picture one could instead conceive the possibility that the collapsing matter
occupies (in the quantum mechanical sense of the Compton length) a large volume inside the black hole
and gives rise to an effective gravitational potential that differs significantly from the General Relativistic description.
This is precisely the reason we developed the ``bootstrapped'' Newtonian picture of a homogeneous
source, not to be taken as a model of phenomenological relevance for compact objects like neutron stars,
but as a toy model of gravity tailored to further investigating this quantum picture of black holes.
It is indeed rather likely that, in order to describe more accurately astrophysical compact objects,
one would need to add more interacting terms to the Lagrangian~\eqref{LagrV}, so that the gravitational
potential outside the gravitational radius approaches the usual post-Newtonian expansion of the
General Relativistic Schwarzschild metric.
\par
Although the explicit quantum investigation of the bootstrapped Newtonian ball is left for a future work,
we can here highlight a few relevant features.
First of all, we have already noted that the (exact) vacuum solution $V_{\rm c}$ in Eq.~\eqref{sol0} tracks
the Newtonian behaviour, and its derivative therefore gives an attractive force for all (finite) values of $r>0$.
This is in clear contrast with the weak field expansion in Eq.~\eqref{Vlarge}, which instead provides a
repulsive force for $r\le \Rh=2\,\gn\,M$, if one only includes the first post-Newtonian term $V_{\rm PN}$.
In particular, one could apply a Newtonian argument to $V_{\rm c}$ and define the ``horizon'' as the
place where the escape velocity of a particle subjected to it would equal the speed of light.
This occurs for $2\,V_{\rm c}(\rh)=-1$, which yields
\be
\rh
=
\frac{6\,\gn\,M}{3\,\sqrt{3}-1}
\simeq
1.4\,\gn\,M
\ ,
\ee
or $\rh\simeq 0.7\,\Rh$. 
We remark once more that a source of size $R$ this small could not be studied in the weak field
expansion to first post-Newtonian order, as that approximation does not hold for this range
of the radial coordinate.
In fact, it would also violate the Buchdahl limit~\eqref{Rbl} and could not be a stable configuration
according to General Relativity.
\par
Since black holes in General Relativity are regions of empty space with singular sources in the very
centre, we can consider the case of a source with radius $R\ll\rh$.
Let us note that, at least for a homogeneous source, we always have $M>M_0$ and
Eq.~\eqref{MvM0sf} then implies that the ratio
\be
\frac{M}{M_0}
\simeq
\left(\frac{6\,\gn\,M_0}{R}\right)^{1/2}
\simeq
\left(\frac{6\,\gn\,M}{R}\right)^{1/3}
\gg
1
\ ,
\label{M0vMsf}
\ee
for $R\ll\rh$.
The area law of black holes~\cite{bekenstein} states that the mass $M$ should measure
the number of gravitational degrees of freedom, which is also confirmed explicitly by the number of
gravitons~\eqref{Ng} in the corpuscular picture~\cite{DvaliGomez}.
The above Eq.~\eqref{M0vMsf} therefore appears in line with the expectation that the number of gravitational
degrees of freedom should largely overcome the number of matter degrees of freedom
(proportional to $M_0$) in a black hole.
\par
We have already argued in Section~\ref{ss:pnpressure} why the mass parameter $M$ of the outer vacuum
solution~\eqref{sol0} should equal the total energy $E$ of the system.
Moreover, the previous considerations naturally lead us to look more closely at the relation between $M$
and the proper mass $M_0$.
In particular, if we add the proper energy $M_0$ to the gravitational potential energy $U_{\rm G}$ from
Eq.~\eqref{Uapp}, we find
\be
E_{0{\rm G}}
\equiv
M_0+U_{\rm G}
\simeq
M_0
\left(1-\alpha\,\frac{\gn\,M_0}{R}-\beta\,\frac{\gn^2\,M_0^2}{R^2}\right)
\ ,
\label{Eapp}
\ee
where $\alpha$ and $\beta$ are numerical coefficients of order one, which can be estimated
using the two approximations employed in Eq.~\eqref{Uapp} or computed numerically.
We then find that the energy $E_{0{\rm G}}$ is positive for low compactness sources with $\gn\, M_0\ll R$,
it vanishes for $\gn\,M_0\simeq 0.6\,R$ and becomes negative for very compact sources with $\gn\, M_0\gg R$.
For $R\simeq \rh$, or
\be
\gn\,M
\simeq
0.7\,\rh
\ ,
\ee
one can solve the matching conditions~\eqref{cont} numerically and finds
\be
\rh
\simeq
2\,{\gn\,M_0}
\quad
{\rm and}
\quad
C
\simeq
1.2
\ .
\ee
Moreover, this value of $M_0$ is also close to the case that makes the energy~\eqref{Eapp},
\be
E_{0{\rm G}}(\rh)=M_0+U_{\rm G}(\rh)\simeq 0
\ .
\label{maxp}
\ee
We shall thus conclude by noting that the vanishing of $E_{0{\rm G}}$ at (the threshold of) black hole formation
appears as another form of the ``maximal packing'' condition at the heart of the corpuscular
picture~\cite{DvaliGomez,Casadio:2016zpl,Casadio:2017cdv}.
In fact, Eq.~\eqref{maxp} is very close to the form of the maximal packing condition that is implemented
in the quantum harmonic model of corpuscular black holes~\cite{QHBH}.
Whether this maximal packing conditions implies that the baryonic matter is completely delocalised inside the
horizon like the gravitons requires a fully quantum treatment of the system and is left for future investigations. 
\section*{Acknowledgments}
The authors would like to thank M.F.~de~Laurentis, A.~Giugno, A.~Giusti, and J.~Mureika
for useful comments and suggestions.
R.C.~and M.L.~are partially supported by the INFN grant FLAG.
The work of R.C.~has also been carried out in the framework
of activities of the National Group of Mathematical Physics (GNFM, INdAM)
and COST action {\em Cantata\/}. 
O.M.~is supported by the grant Laplas~V of the Romanian National Authority for Scientific
Research. 
\appendix
\section{Potential energy}
\label{A:potential}
\setcounter{equation}{0}
The gravitational potential energy of the vacuum solution~\eqref{sol0} can be easily computed from
the effective Lagrangian~\eqref{LagrV} with $\rho=0$ and reads
\be
U_{\rm G}^+[V_{\rm c}]
=
-L^+[V_{\rm c}]
&\!\!=\!\!&
\frac{1}{2\,\gn}
\int_R^\infty
r^2\,\d r
\left(1-4\,V\right)
\left(V'\right)^2
\nonumber
\\
&\!\!=\!\!&
\frac{\gn\,M^2}{2}
\int_R^\infty
\frac{\d r}{r^2}
=
\frac{\gn\,M^2}{2\,R}
\ ,
\label{UballOUT}
\ee
where $R$ can be simply viewed as a cut-off to regularise the result, in analogy with the Newtonian
case.
\par
If we instead consider a homogeneous ball in vacuum, the length $R$ naturally becomes the radius
of the source.
We again switch to dimensionless units and the total gravitational potential energy for the Newtonian
solution~\eqref{Vn}
is found to be
\be
\tilde U_{\rm G}[\tilde V_{\rm N}]
&\!\!=\!\!&
-\tilde L[\tilde V_{\rm N}]
=
\frac{1}{2}
\int_0^\infty
\tilde r^2\,\d\tilde  r
\left[
\left(1-4\,\tilde V_{\rm N}\right)
\left(\tilde V_{\rm N}'\right)^2
+8\,\pi\,\tilde \rho\,\tilde V_{\rm N}\left(1-2\,\tilde V_{\rm N}\right)
\right]
\nonumber
\\
&\!\!=\!\!&
3\,\tilde M_0
\int_0^1
\tilde r^2\,\d\tilde  r\,
\tilde V_{\rm N}
\left(
1-
2\,\tilde V_{\rm N}^2
\right)
+
\frac{1}{2}
\int_0^\infty
\tilde r^2\,\d\tilde  r
\left[
\left(1-4\,\tilde V_{\rm N}\right)
\left(\tilde V_{\rm N}'\right)^2
\right]
\nonumber
\\
&\!\!=\!\!&
-\frac{3\,\tilde M_0^2}{5}
+\frac{51\,\tilde M_0^3}{35}
=
-\frac{3\,\tilde M_0^2}{5}
\left(1-\frac{255\,\tilde M_0}{105}\right)
\ ,
\label{UballN}
\ee
in which we recognize the standard (negative) Newtonian contribution obtained from the first line in Eq.~\eqref{Unn},
that is
\be 
\tilde U_{\rm N}
=
\frac{\tilde M_0^2}{4}
\int_0^1
{\tilde r}^2\,\d {\tilde  r}\,
\tilde V_{\rm N}
=
-
\frac{3\,\tilde M_0^2}{5}
\ .
\label{UN}
\ee
Like for the exact vacuum solution~\eqref{sol0}, the last integral in the second line of Eq.~\eqref{UballN}
yields a positive contribution,
which is however overcome by the negative Newtonian energy and its (still negative) correction
arising where $\tilde\rho\not = 0$.
\par
Given the correct solution to the effective equation~\eqref{tEOMV} is very close to the Newtonian expression~\eqref{Vn-}
inside the ball, we can estimate the contribution to the potential energy from the region inside the source as~\footnote{We
also checked that using the approximation~\eqref{Va} with Eq.~\eqref{W} does not alter the result significantly.} 
\be
\tilde U_{\rm G}^-[\tilde V_{\rm c}]
\simeq
-L^-[\tilde V_{\rm N}]
&\!\!=\!\!&
\frac{1}{2}
\int_0^1
\tilde r^2\,\d\tilde  r
\left[
\left(1-4\,\tilde V_{\rm N}\right)
\left(\tilde V_{\rm N}'\right)^2
+6\,\tilde M_0\,\tilde V_{\rm N}
\left(1-2\,\tilde V_{\rm N}\right)
\right]
\nonumber
\\
&\!\!\simeq\!\!&
-\frac{3\,\tilde M_0^2}{2}
\left[
C-\frac{4}{15}
+\tilde M_0
\left(\frac{5}{21}-\frac{24}{15}\,C + 3\, C^2\right)
\right]
\ .
\label{UballIN}
\ee
We can finally estimate the total gravitational energy of the homogeneous ball by adding together
Eq.~\eqref{UballOUT} for the outer vacuum and the contribution~\eqref{UballIN} from the interior,
which yields
\be
\tilde U_{\rm G}
\simeq
\tilde U_{\rm G}^-[\tilde V_{\rm N}]
+
\tilde U_{\rm G}^+[\tilde V_{\rm c}]
\simeq
\left\{
\begin{array}{lcl}
-\strut\displaystyle\frac{3\ \tilde M_0^2}{5}
\left(1+\frac{109\,\tilde M_0}{42}\right)
&
{\rm for}
&
\tilde M_0\ll 1
\\
\\
-{2.22\,\tilde M_0^3}
-{1.61\,\tilde M_0^2}
&
{\rm for}
&
\tilde M_0\gg 1
\ ,
\end{array}
\right.
\label{Uapp}
\ee
in which we used Eqs.~\eqref{MvM0wf} and \eqref{CvM0wf} for $\tilde M_0\ll 1$ and
Eqs.~\eqref{MvM0sf} and \eqref{Csf} for $\tilde M_0\gg 1$.
\begin{figure}[t]
\centering
\includegraphics[width=10cm]{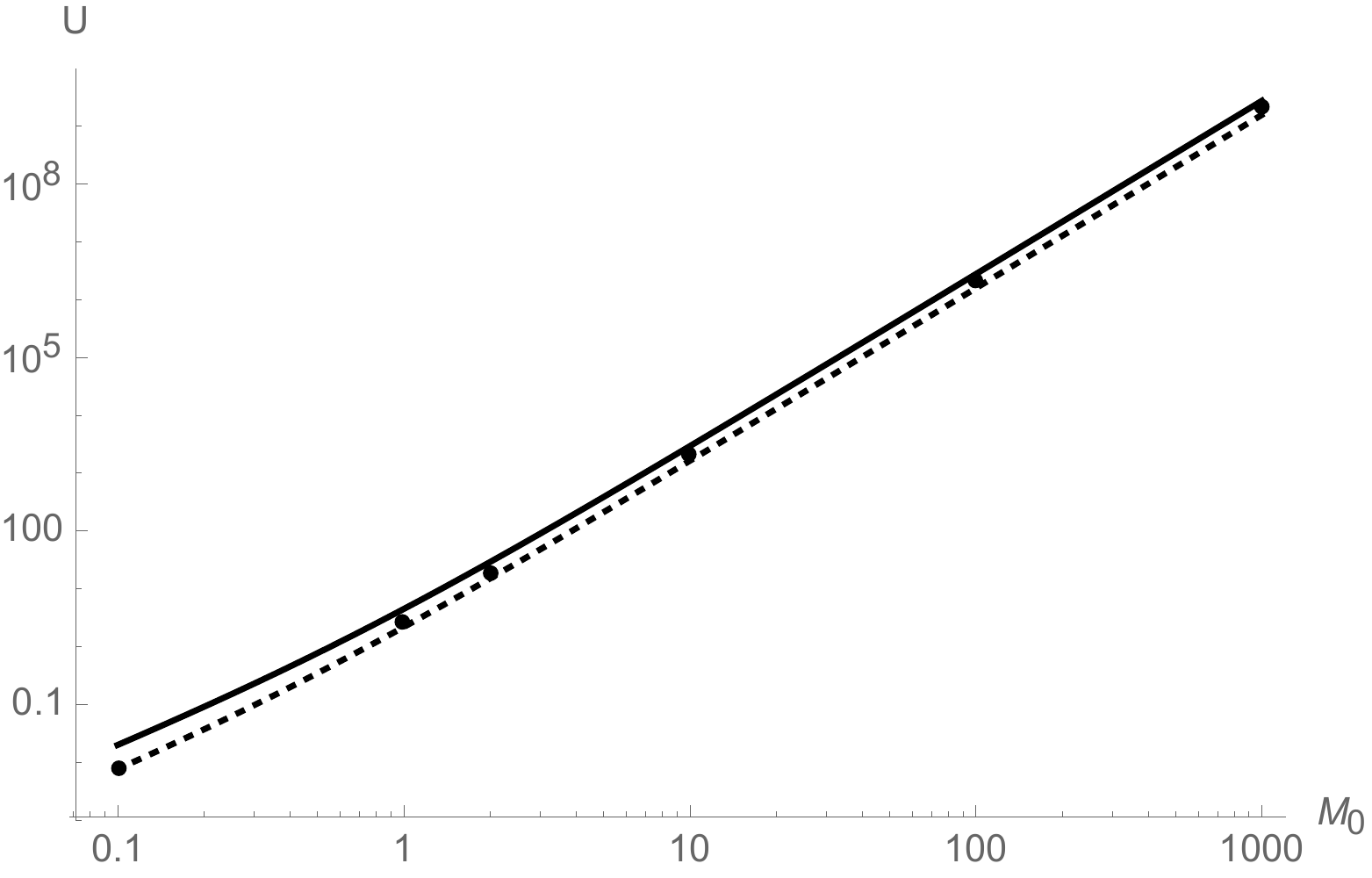}
\caption{Gravitational potential energy $\tilde U_{\rm G}$ evaluated numerically (dots) {\em vs\/} the
analytical approximation~\eqref{Uapp} for $\tilde M_0\gg 1$ (solid line) {\em vs\/} the
analytical approximation~\eqref{Uapp} for $\tilde M_0\ll 1$ (dotted line).}
\label{pU}
\end{figure}
\par
The gravitational energy $\tilde U_{\rm G}$ can also be computed numerically for given values of
$\tilde M_0$, and compared with the above analytical approximations.
From the plot in Fig.~\ref{pU}, it appears that both approximate expressions reproduce the
numerical results fairly well. 
\section{Junction conditions}
\label{A:junction}
\setcounter{equation}{0}
Because the shift symmetry is lost, we cannot just match the outer potential~\eqref{sol0}
with just any interior.
It is then useful to analyse the matching conditions at $r=R$.
This can be done as usual by integrating Eq.~\eqref{EOMV} on a shell of thickness $0<2\,\epsilon\ll R$
around $r=R$,
\be
0
=
\int_{R-\epsilon}^{R+\epsilon}
r^2\,\d r
\left[
\left(1-4\,V\right)\left(\triangle V-4\,\pi\,\gn\,\rho\right)
-
2\left(V'\right)^2
\right]
\equiv
I_1+I_2+I_3
\ .
\ee
For the first term we obtain
\be
I_1
&\!\!=\!\!&
\int_{R-\epsilon}^{R+\epsilon}
\d r
\left(1-4\,V\right)\left(r^2\,V'\right)'
\nonumber
\\
&\!\!=\!\!&
\left[\left(1-4\,V\right)\left(r^2\,V'\right)\right]_{R-\epsilon}^{R+\epsilon}
+4
\int_{R-\epsilon}^{R+\epsilon}
r^2\,\d r
\left(V'\right)^2
\ ,
\ee
so that
\be
I_1+I_3
&\!\!=\!\!&
\left[\left(1-4\,V\right)\left(r^2\,V'\right)\right]_{R-\epsilon}^{R+\epsilon}
+2
\int_{R-\epsilon}^{R+\epsilon}
r^2\,\d r
\left(V'\right)^2
\ .
\ee
If we assume $\rho$ and $V$ do not diverge at $r=R$, the second term vanishes
for $\epsilon\to 0$,
\be
I_2
=
4\,\pi\,\gn
\int_{R-\epsilon}^{R+\epsilon}
r^2\,\d r
\left(1-4\,V\right)
\rho
\underset{\epsilon\to 0}{\to}
0
\ ,
\ee
and one is left with
\be
0
=
\left[\left(1-4\,V\right)\left(r^2\,V'\right)\right]_{R-\epsilon}^{R+\epsilon}
+2
\int_{R-\epsilon}^{R+\epsilon}
r^2\,\d r
\left(V'\right)^2
\ ,
\ee
which is satisfied for $\epsilon\to 0$ if $V'$ is continuous across $r=R$.
\section{Anti-de~Sitter potential?}
\setcounter{equation}{0}
\label{A:homo}
Eq.~\eqref{EOMV} with the homogeneous density in Eq.~\eqref{HomDens} is solved exactly for $0\le r<R$
by
\be
V_{\rm c}
=
\frac{1}{4}
+
\frac{3\,\gn\,M_0}{8\,R^3}\,r^2
\ ,
\ee
which is clearly positive everywhere, and cannot be matched with a negative outer potential
at $r=R$.
This solution could however be considered as a ``cosmological'' solution similar to the anti-de~Sitter
space.
\section{Comparison with TOV}
\setcounter{equation}{0}
\label{TOV}
It is instructive to compare our result~\eqref{pp} for the pressure with the expectation obtained by solving the
TOV equation for a homogeneous source.
The standard TOV equation relating the pressure and energy density reads~\cite{stephani}
\be
\frac{\d p(r)}{\d r} 
=
\frac{\left[\rho + p(r)\right] \gn
\left[2\, m(r) + 8\, \pi\, p(r)\, r^3\right]}
{2\, r\, \left[2\,\gn \, m(r) - r \right]}
\ ,
\label{tov}
\ee
where $m(r)$ is the mass function
\be
m(r) 
=
4\,\pi \int_{0}^{r} \d r'\,\rho(r')\, r'^{ 2} 
\ .
\label{m(r)}
\ee
Its solution for the homogeneous density~\eqref{HomDens} can be found in exact form
after requiring that $p_{\mathrm{TOV}}(R)=0$.
This yields
\be
\label{ptov}
p_{\mathrm{TOV}}(r)  
&\!\!=\!\!&
\rho(r)\, \frac{\sqrt{R^3  - {2\, \gn\, M_0\, r^2}} 
-
R\,\sqrt{R - { 2\, \gn\, M_0}} }
{3\,R\,\sqrt{R - { 2\, \gn\, M_0}} 
- 
\sqrt{R^3 -  {2\, \gn\, M_0\, r^2}} }
\nonumber
\\
&\!\!=\!\!&
\frac{3\, M_0 \left( \sqrt{R^3  - {2\, \gn\, M_0\, r^2}} 
-
R\,\sqrt{R - { 2\, \gn\, M_0}} \right) }
{4\, \pi\,R^3 \left( 3\,R\,\sqrt{R - { 2\, \gn\, M_0}} 
- 
\sqrt{R^3 -  {2\, \gn\, M_0\, r^2}} \right)}
\ .
\ee
\par
This TOV pressure can easily be compared to our Post-Newtonian result from Eq.~\eqref{pp} in the small
compactness regime ${\gn\,M_0}\ll R$, that is far from the Buchdahl limit. 
Considering the first two leading terms, the TOV pressure is approximately equal to 
\be
\label{ptov1}
p_{\mathrm{TOV}}(r)  
\simeq
\frac{3\left(R^2-r^2\right)\gn\,M_0^2}{8\,\pi\,R^6}
\left(
1
+
\frac{8\,\gn\,M_0}{3\,R}
\right)
\ ,
\ee
while the Post-Newtonian pressure from Eq.~\eqref{pp} is approximately
\be
\label{pp1}
p(r)  
\simeq
\frac{3\left(R^2-r^2\right)\gn\,M_0^2}{8\,\pi\,R^6}
\left[
1
+
\frac{\left(R^2+r^2\right)\gn\,M_0}{5\,R^3}
\right]
\ . 
\ee
Both the lowest order term of the TOV pressure and the one of the Post-Newtonian pressure
are equal to what one calculates in the Newtonian case~\eqref{pN}. We can also remark that the
next-to-leading order contributions are much smaller.  
Therefore, as expected, in this limit the pressure inside the objects obtained in our model can be well
approximated by the Newtonian pressure. \par
\begin{figure}[t]
\centering
\includegraphics[width=8cm]{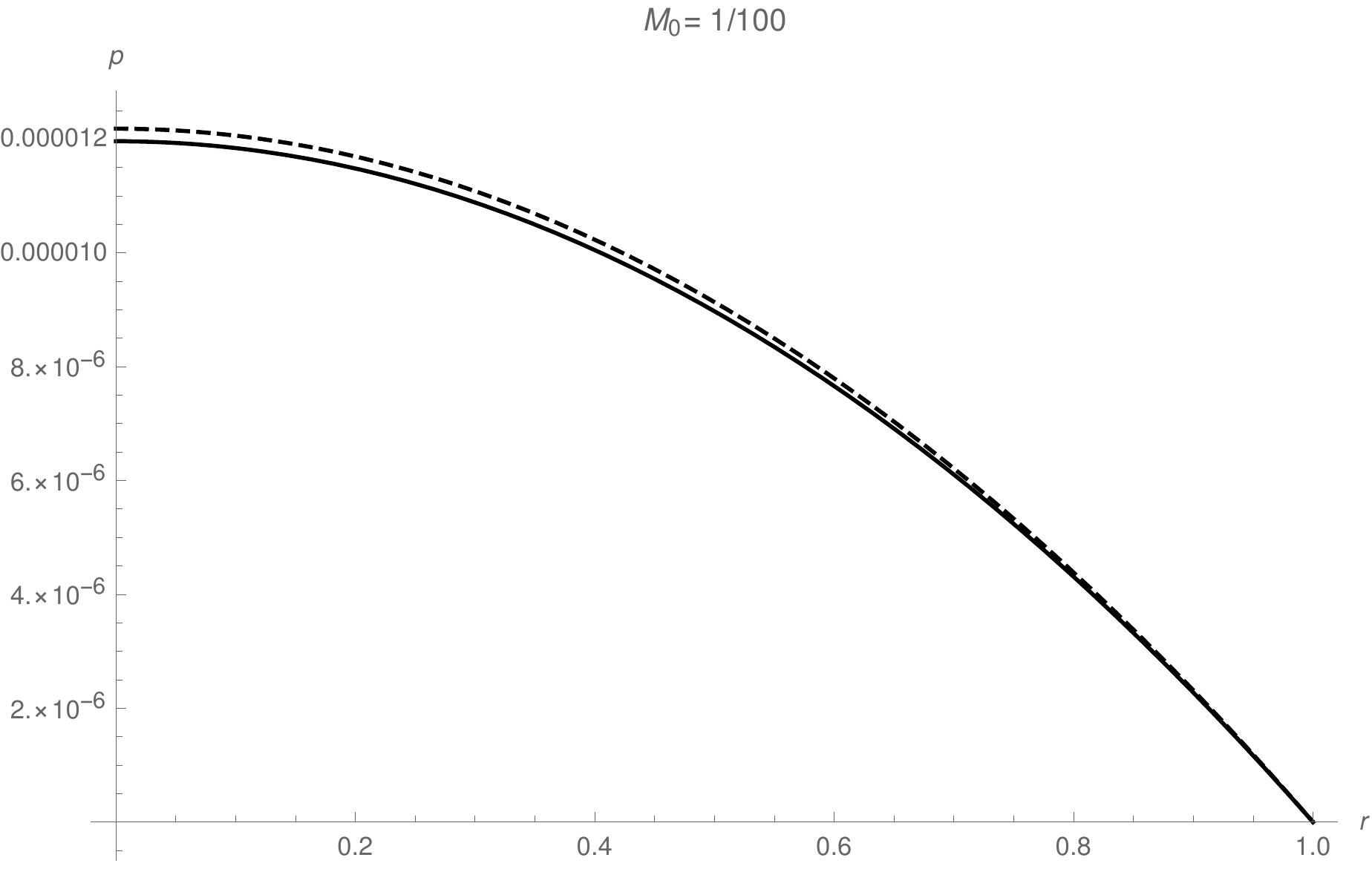}
$\ $
\includegraphics[width=8cm]{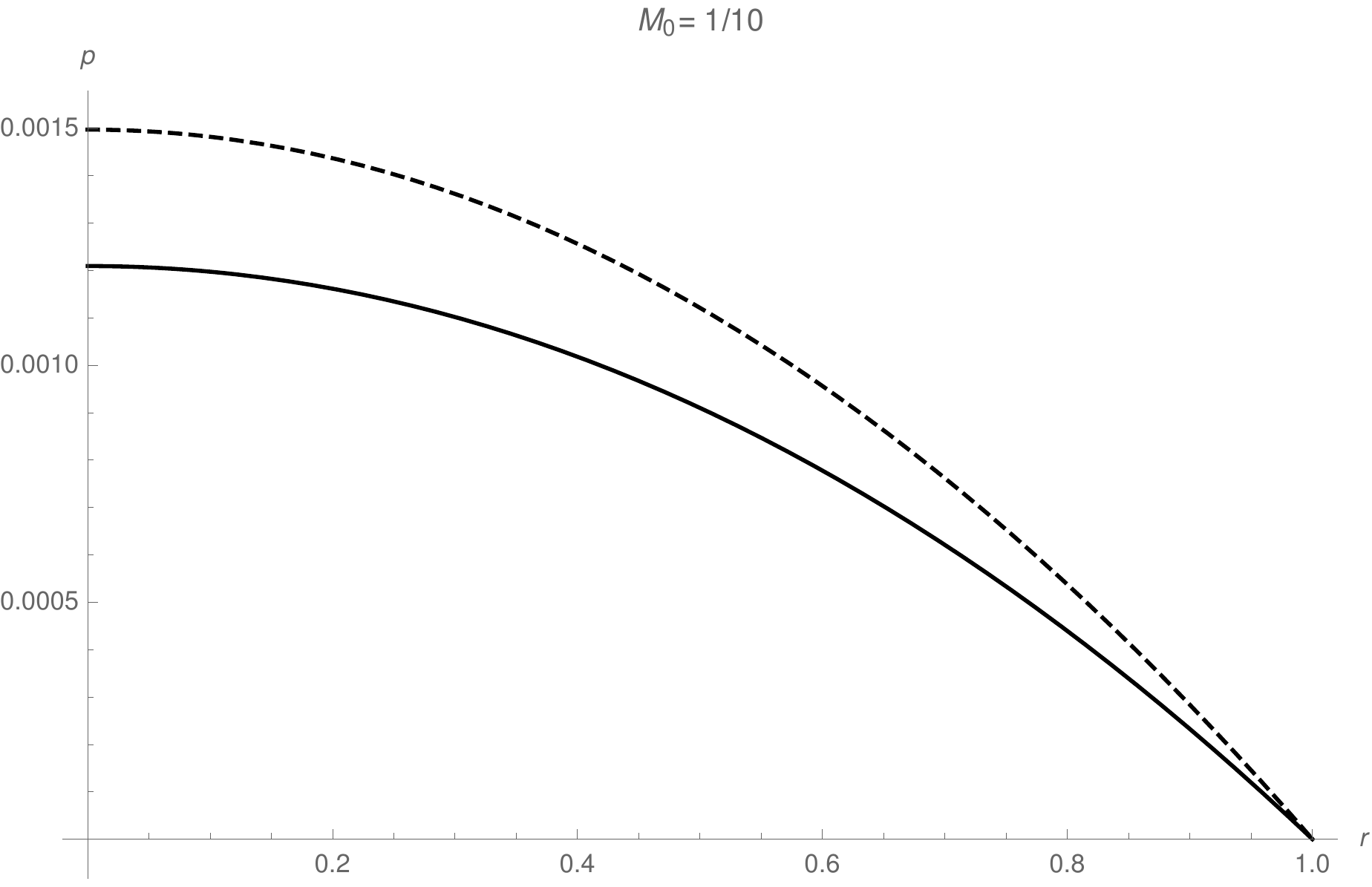}
\caption{Numerical solution (solid line) {\em vs\/} $p_{\mathrm{TOV}}$ (dashed line) for $\tilde M_0=1/100$ (left panel)
and $\tilde M_0=1/10$ (right panel).}
\label{PTOV}
\end{figure}

A comparison between the TOV solution and the numerical evaluation of the pressure within our model
is displayed in Fig.~\ref{PTOV}. The plots also show that the two are in good agreement as long as
the source is sufficiently less compact than the Buchdahl limit~\eqref{Rbl}.
This is not surprising, since the pressure~\eqref{pp} remains finite for any $0\le r < R$,
whereas $p_{\mathrm{TOV}}(0)$ in Eq.~\eqref{ptov} diverges for $R\to R_{\rm BL}^+$.
\end{document}